\documentclass[prd,aps,floats,floatfix,superscriptaddress,preprintnumbers,showpacs,eqsecnum,nofootinbib,twocolumn]{revtex4}
\usepackage{amsfonts,amsmath,epsfig,amssymb,latexsym,color,graphicx,array,subfigure,bm,float }

\definecolor{red}{rgb}{1,0,0}
\definecolor{blue }{rgb}{0,0,1}
\definecolor{green}{rgb}{0,1,0}


\newcommand{\bea}{\begin{eqnarray}}
\newcommand{\ena}{\end{eqnarray}}
\newcommand{\beq}{\begin{eqnarray}}
\newcommand{\eeq}{\end{eqnarray}}
\newcommand{\be}{\begin{equation}}
\newcommand{\ee}{\end{equation}}
\newcommand{\beann}{\begin{eqnarray*}}
\newcommand{\enann}{\end{eqnarray*}}

\newcommand{\dsl}{\pa \kern-0.5em /}

\newcommand{\pa}{\partial}

\newcommand{\nn}{\nonumber\\}

\newcommand{\vect}[1]{\,\!\!\!\mbox{ \boldmath $#1$}}

\begin{document}

\date{\today}

\title{Einstein Cluster as Central Spiky Distribution of Galactic Dark Matter}

\author{Kei-ichi Maeda}
\affiliation{
Department of Physics, Waseda University, 3-4-1 Okubo, Shinjuku, Tokyo 169-8555, Japan}
\affiliation{Center for Gravitational Physics and Quantum Information, Yukawa Institute for Theoretical Physics, Kyoto University, 606-8502, Kyoto, Japan}

\author{Vitor Cardoso}
\affiliation{CENTRA, Departamento de F\'{\i}sica, Instituto Superior T\'ecnico -- IST, Universidade de Lisboa -- UL,
Avenida Rovisco Pais 1, 1049 Lisboa, Portugal and}
\affiliation{Niels Bohr International Academy, Niels Bohr Institute, Blegdamsvej 17, 2100 Copenhagen, Denmark}

\author{Anzhong Wang}
\affiliation{GCAP-CASPER, Physics Department, Baylor University, Waco, Texas 76798-7316, US}

\begin{abstract}
Using the Einstein cluster models, we construct a fully relativistic, spherically symmetric, spiky structure of matter distribution near a supermassive black hole. We introduce and discuss three simple toy models, together with a more realistic model, which includes a Hernquist-type distribution with a typical galaxy scale. We find that the innermost stable circular orbit (ISCO) depends on the details of the environment, and lies between the photon radius (at $3M_{\rm BH}$) and the ISCO radius of an isolated black hole of mass $M_{\rm BH}$ (at $6M_{\rm BH}$).
\end{abstract}

\maketitle




\section{Introduction}
The nature of dark matter (DM) and its role in standard model of particle physics is one of the most mysterious problems in modern cosmology. We know it exists via its gravitational effects, but we are ignorant of its properties, including mass or couplings to other fields. The range of possibilities is extremely large: DM could be a new fundamental particle, or it could simply be a large amount of tiny primordial black holes (PBHs), or a plethora of new fields, among other possibilities. Despite the numerous searches for specific DM candidates, no compelling candidate has yet been found~\cite{Freese:2008cz,Navarro:1995iw,Clowe:2006eq,Bertone:2004pz,Kahlhoefer:2017dnp,PerezdelosHeros:2020qyt}.

One possible approach to understand the nature and local distribution of DM consists on studying
specific events near a supermassive black hole (SMBH) or intermediate mass black hole (IMBH).
The annihilation of DM particles and consequent sourcing of high luminosity events has been discussed 
as a possible smoking gun of DM~\cite{Gondolo:1999ef}. A powerful alternative consists on the observation of gravitational waves, which is to a large extent independent of the strength of the coupling between DM and the standard model of particle physics~\cite{PhysRevD.91.044045, Cardoso:2021wlq,Duque:2023cac}. In addition, precise monitoring of the 
periapsis shift and overall motion of stars nearby BHs could also 
reveal some features of DM distribution \cite{2008ApJ...689.1044G,2019A&A...625L..10G,GRAVITY:2021xju,Igata:2022rcm}.

Most DM distributions are studied in a fixed, flat-space background. One of the most popular distributions is the Navarro–Frenk–White (NFW) profile, which is consistent with observations and $N$-body simulations. Other profiles such as Hernquist-type distribution have also been proposed~\cite{1990ApJ...356..359H}. 
The adiabatic growth of a BH 
may induce a spiky, dense structure near BHs~\cite{Gondolo:1999ef}. The relativistic extension of the spiky structure near 
SMBH was given by~\cite{Ferrer:2017xwm}.
One of the striking conclusions of these works is that the inner boundary of the spiky DM distribution 
is below the innermost stable circular orbit (ISCO) radius; in particular, the inner boundary is located at an areal radius $4M_{\rm BH}$ for a spherically symmetric system, with $M_{\rm BH}$ the BH mass.

The DM distribution close to BHs is particularly interesting: its density in the strong gravity regime can be considerably large, and hence effects on gravitational wave emission can be important. Thus, in the context of observations concerning events in highly dynamical, strong-field gravity, a careful modelling of DM distribution in a fully general-relativistic setting is required. 
Especially, when we take into account the self-gravity of the DM distribution, the curved spacetime description is no longer that of a vacuum Kerr geometry.
Fully general relativistic solutions describing spherically symmetric BHs immersed in environments were discussed recently~\cite{Cardoso:2021wlq} assuming an Hernquist-type DM distribution.
The DM is distributed all throughout the BH exterior regions in the original work, violating the dominant energy condition is a region close to the horizon. A simple modification of the original profile eliminates this issue~\cite{Speeney:2024mas}; other analytic form of DM distribution was also given in Ref.~\cite{Shen:2024qxv,Shen:2023erj}.

In this note, we provide simple analytic models with spiky DM distribution within the so-called Einstein cluster construction. In fact, although we have DM in mind, our results apply as well for standard, baryonic matter environments. There are several works on DM distribution by use of the Einstein cluster~\cite{10.1111/j.1365-2966.2007.11977.x, Jusufi2023,acharyya2023modellingeinsteinclusterusing}. Here we focus on the possibility that some distributions may yield smaller ISCO radius, below 
the ISCO radius of the central BH ($6M_{\rm BH}$).

After we introduce the basic equations for the Einstein cluster in \S \ref{Einstein_cluster}, we discuss  on how to find the ISCO radius of the DM distribution when we take into account self-gravity of DM in \S \ref{ISCO_radius}.  We then present  three toy models
with the ISCO radius smaller than $6 M_{\rm BH}$ in \S \ref{toy_model}.
In \S \ref{realistic_model}, we present a slightly realistic models with a typical galactic scale where the distribution profile changes.
The summary and discussion follow in \S \ref{summary}.

\section{A spherically symmetric Einstein cluster spacetime}
\label{Einstein_cluster}
We consider a spherically symmetric spacetime, described by the line element
\bea
ds^2=-f(r) dt^2+\frac{dr^2}{1-\frac{2m(r)}{r}}+r^2d\Omega^2
\,.
\label{spherical_metric}
\ena
In vacuum, the general solution is the Schwarzschild spacetime,
\beann
m(r)=M_{\rm BH}\,,~~f(r)=f_0\left(1-{2M_{\rm BH}\over r}\right)
\,,
\enann
where $M_{\rm BH}$ is the BH mass and $f_0$ is an integration constant.

In this note, we consider DM distribution outside the
 event horizon. 
We assume that DM is composed of particles 
moving along circular geodesics just as the Einstein cluster. Since the radial component of the pressure is assumed to be zero ($P_r=0$) as we will show it explicitly for the present model,
the Einstein equations and energy-momentum conservation give 
 \bea
 m'&=&4\pi r^2 \rho\,,
 \label{Eeq1}
 \\
 {f'\over f}&=&{2m(r)\over r(r-2m(r))}\,,
\label{Eeq2}
 \\
 P_t&=&{m(r)\over 2(r-2m(r))}\rho
 \label{em_cons}
\,.
 \ena

For circular orbits, the 4-velocity obeys
\beann
u_0&=&-E\,,~u^0\,=\,{E\over f}\,,\\
u_\phi&=&L_z\,,~u^\phi\,=\,{L_z\over r^2\sin^2\theta}\,,
\\
r^4 (u^\theta)^2&=&L^2-{L_z^2\over \sin^2\theta}
\,.
\enann
With the normalization $u^\mu u_\mu=-1$,
we find the radial equation
\beann
{f(r)\over 1-{2m(r)\over r}}\left({dr\over d\tau}\right)^2=E^2-V_{\rm eff}^2(r;L)
\,,
\enann
where the effective potential $V_{\rm eff}^2$ is defined by
\bea
V_{\rm eff}^2=f(r)\left(1+{L^2\over r^2}\right)\,.
\label{eff_potential}
\ena

The radius of a circular orbit is given by the equation
\bea
{dV_{\rm eff}^2\over dr}={2f\over r^3(r-2m(r))}\left[m(r)r^2-L^2(r-3m(r))\right]=0
\,,
\nn
~
\label{circular_radius}
\ena
where we used Eq. (\ref{Eeq2}).

The energy $E$ and angular momentum $\vect{L}\, (L=|\vect{L}|\,, L_z)$
of a particle on a circular orbit in the spacetime (\ref{spherical_metric}) 
 are given by 
\bea
E^2&=&f(r)\left(1+{L^2\over r^2}\right)={r-2m(r)\over r-3m(r)}f(r)\,,
\label{energy_circular}
\\
L_z&=&L\cos\theta\,,
\\
L^2&=&{m(r)r^2\over r-3m(r)}
\label{ang_circular}
\,.
\ena
Note that since $m(r)>0$ and $f>0$,
we find that $r>3m(r)$ from the above equations, which means that
photon sphere in the present model  does not exist inside the 
dark DM distribution.

The energy-momentum tensor of $N$ particles is given by 
\beann
T^\mu_{~\nu}=\mu_0 \sum_{I=1}^N \int d\tau_{\rm I} { u_{\rm I}^\mu {u_{\rm I}}_\nu \over \sqrt{-g}}
 \delta(x-z_{\rm I})
 \,,
\enann
where $\mu_0$ is a particle mass and $u_{\rm I}^\mu$ is 4-velocity of the $I$-th particle 
 in the spacetime given by the metric (\ref{spherical_metric}).
The energy density and pressures are given by

\begin{widetext}
\beann
\rho(\vect{r})&=&u^0 u_0 n(\vect{r})={E^2\over f}n(\vect{r})=\left(1+{L^2\over r^2}\right)n(\vect{r})={r-2m(r)\over r-3m(r)}n(\vect{r})\,,
\\
P_r(\vect{r})&=&0\,,
\\
P_t(\vect{r})&=&\langle u^\phi u_\phi\rangle  n(\vect{r})={L^2\langle \cos^2\theta\rangle \over r^2}n(\vect{r})
={L^2\over 2 r^2}n(\vect{r})={m(r)\over 2(r-3m(r))}n(\vect{r})\,,
\enann
where $n(\vect{r})$ is the unknown distribution function of particles, determined by the energy density $\rho(\vect{r})$.

 \section{ISCO radius and Conditions near ISCO radius}
 \label{ISCO_radius}

We then look for the radius of the ISCO in the spacetime (\ref{spherical_metric}), which is obtained 
by Eq.(\ref{circular_radius}) and 
%
\be
{d^2V_{\rm eff}^2\over dr^2}
={2f\over r^4(r-2m(r))^2}
\times \left[
m'r^4-2mr^2(r-2m)+L^2\left(m'r^2+3(r-2m)(r-4m)\right)\right]=0\,,
\label{inflection_point}
\ee
\end{widetext}
where we again used Eq. (\ref{Eeq2}).
Eliminating $L^2$, we find 
the stability condition for the circular orbit
\bea
r^2 m'(r)+r m(r)-6m^2(r) \geq 0
\label{stability}
\,,
\ena
where the equality corresponds to 
the inflection point.
Hence solving the equation
\bea
r^2 m'(r)+r m(r)-6m^2(r)=0
\label{eq_inflection}
\,,
\ena
we obtain the ISCO radius $r_{\rm I}$ as the minimum value of the possible solutions.
We then find the energy $E_{\rm ISCO}$ and angular momentum 
$L_{\rm ISCO}$ of a particle at 
the ISCO radius 
from Eqs. (\ref{energy_circular}) and (\ref{ang_circular}) by inserting $r=r_{\rm I}$.

Now we consider the behaviour of metric and matter distribution near the ISCO radius. 
We assume there  is no matter fluid below the ISCO radius $r_{\rm I}$.
We then assume mass function $m(r)$ near the ISCO radius as 
\beann
m(r)=M_{\rm BH}\left[1+\alpha(r-r_{\rm I})^p/M_{\rm BH}^p\right]~~(r\geq r_{\rm I})
\,,
\enann
where $\alpha>0$. There is no solution for $p<0$.
In the case of $p=0$, we have to put a mass shell at $ r_{\rm I}$ and $r_{\rm I}=6m( r_{\rm I})=6(M_{\rm BH}+\alpha)$, which is a trivial case. When $0<p<1$, $m'$ diverges at $r_{\rm I}$ and there is no solution for Eq. (\ref{eq_inflection}).
For $p>1$, we find $r_{\rm I}=6M_{\rm BH}$ because $m'(r_{\rm I})=0$. 
This gives usual BH ISCO radius. 
Note that if the energy density vanishes at the ISCO radius, i.e.,
$\rho\propto (r-r_{\rm I})^{\tilde p} ~(\tilde p>0)$, we find $p=\tilde p +1>1$.

We find a non-trivial solution when $p=1$.
Eq. (\ref{eq_inflection}) becomes
\bea
\alpha r_{\rm I}^2 +M_{\rm BH}r_{\rm I}-6M_{\rm BH}^2=0\,,
\label{alpha_rISCO}
\ena
i.e.,
\bea
m(r)&=&M_{\rm BH}+\alpha(r-r_{\rm I})~~(r\geq r_{\rm I})\,,
\label{mass_fn}
\\
r_{\rm I}&=&{M_{\rm BH}\over 2\alpha}\left(-1+\sqrt{1+24\alpha
}\right)\,.
\label{ISCO_p=1}
\ena

\begin{figure}[ht]
\includegraphics[width=6.5cm]{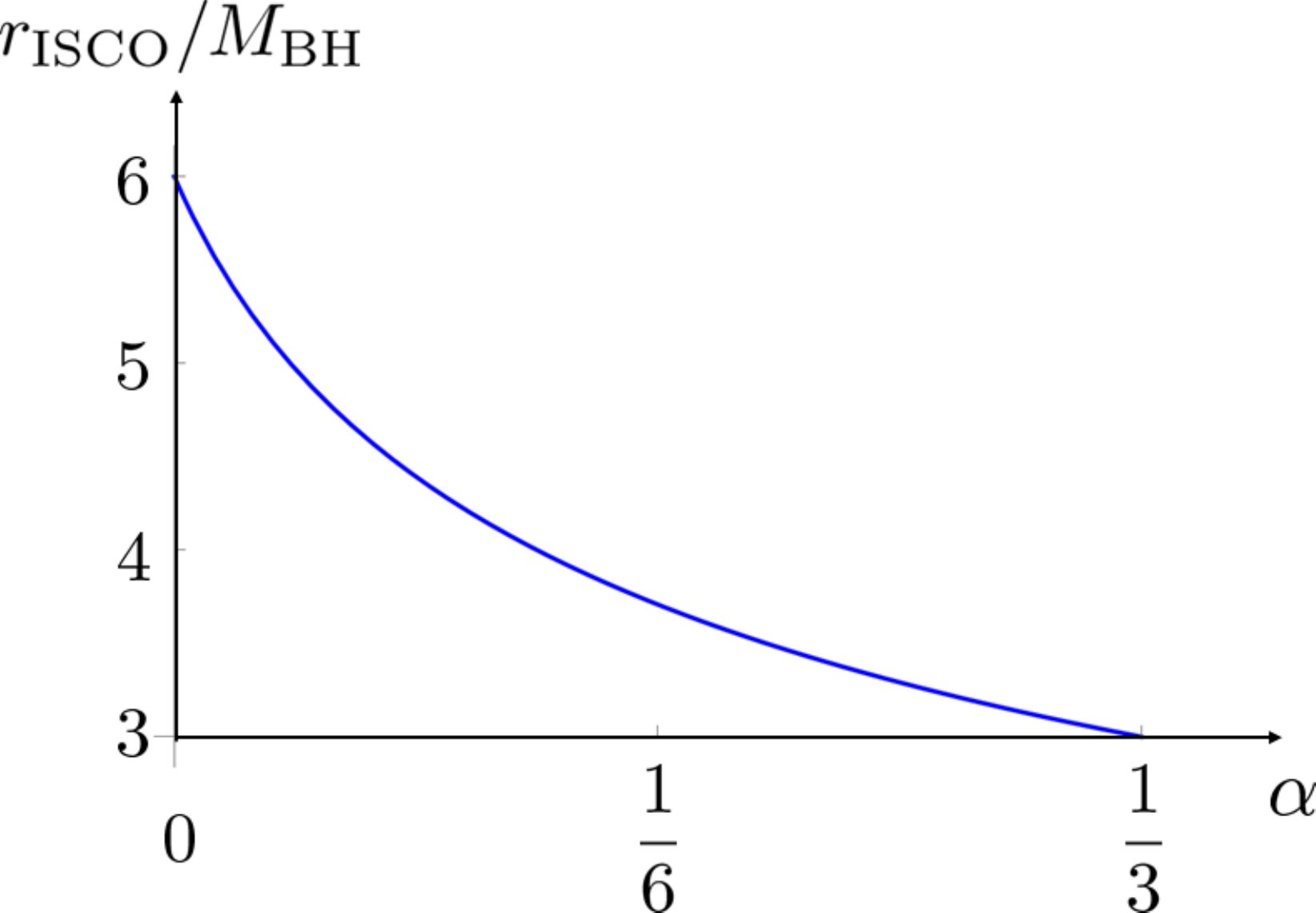}
\caption{The relation between the ISCO radius ($r_{\rm I}$)  and the parameter $\alpha$ for the case of $p=1$.}
\label{ISCO}
\end{figure}
The condition of $r>3m(r)$ gives $\alpha<{1\over 3}$.
We show $r_{\rm I}$ in terms of $\alpha$ in Fig. \ref{ISCO}.
In the limit of $\alpha\rightarrow {1\over 3}$, we find 
$r_{\rm I}\rightarrow 3M_{\rm BH}$, while for $\alpha=0$, we recover 
the vacuum BH ISCO radius ($r_{\rm I}\rightarrow 6M_{\rm BH}$).

\section{Simple Toy  Models}
\label{toy_model}

One interesting question is whether DM can exist  inside the vacuum BH ISCO radius
($<6M_{\rm BH}$). In a realistic DM distribution, 
we have to take into account 
the initial conditions and diffusion process
of particles in a realistic situation. 
However, in this note, we focus solely on the above point concerning existence, and construct simple models with $r_{\rm I}<6M_{\rm BH}$, 
giving mass distribution $m(r)$.

In what follows, we shall normalize all quantities by the BH mass and set 
$M_{\rm BH}=1$.

\subsection{Model I (``Isothermal'' distribution)}
First we assume the (simplistic) mass function
\beann
m(r)=1+\alpha(r-r_{\rm I}) \,,
\enann
where $0<\alpha<1/3$ is a constant.
We find
\beann
\rho={m'\over 4\pi r^2}={\alpha\over 4\pi r^2}
\,,
\enann
an ``isothermal''-like distribution. The metric function $f$ is easily integrated,
\bea
f={c_0\over r}\left(r-r_*\right)^{1\over 1-2\alpha}~~(r\geq r_{\rm I})
\,,
\ena
where 
\beann
r_*\equiv 
{2(1-\alpha r_{\rm I})\over 1-2\alpha}\,,
\enann
and $c_0$ is an integration constant, fixed by the outer boundary conditions.
Stability condition \eqref{stability} is always satisfied. Interestingly, for $\alpha=1/3$ and ISCO radius at $r_{\rm I}=3$ (the photon sphere radius), then every point satisfies the inflection condition (\ref{eq_inflection}). Every point is marginally stable.

Since $m(r)\rightarrow \infty$ as $r\rightarrow \infty$, we  assume that matter distribution is truncated, and exists only between $r_{\rm I}$ and $r_{\rm O}$. The total mass of the system is given by $M=m(r_{\rm O})=1+\alpha(r_{\rm O}-r_{\rm I})$. Outside, the metric is given by the Schwarzschild metric with mass $M$. We can confirm that 
the outer radius $r_{\rm O}$ is always larger than the ISCO radius of the mass $M$, i.e., 
$r_{\rm O}\geq 6M$. The metric is given by
\beann
m(r)&=&\left\{
\begin{array}{cc}
1 & (r<r_{\rm I})
\\
1+\alpha\left(r-r_{\rm I}\right)& (r_{\rm I}<r<r_{\rm O})\,\, ,
\\
M & (r>r_{\rm O})
\\
\end{array}
\right.
\\
f(r)&=&\left\{
\begin{array}{cc}
f_0\left(1-{2\over r} \right)& (r<r_{\rm I})
\\
{c_0\over r}\left(r-r_*\right)^{1\over 1-2\alpha}& (r_{\rm I}<r<r_{\rm O})\,\, ,
\\
1-{2M \over r}& (r>r_{\rm O})
\\
\end{array}
\right.
\enann
where
\beann
f_0&=& {r_{\rm O}-2M\over r_{\rm I}-2}\left({r_{\rm I}-r_*\over r_{\rm O}-r_*}\right)^{1\over 1-2\alpha}\,,
\\
c_0&=&{r_{\rm O}-2M\over (r_{\rm O}-r_*)^{1\over 1-2\alpha}}\,.
\enann
Note that the tangential pressure is given by
\beann
P_t={m(r)\rho(r)\over 2(r-2m(r))}.
\enann
\begin{figure}[ht]
\includegraphics[width=7cm]{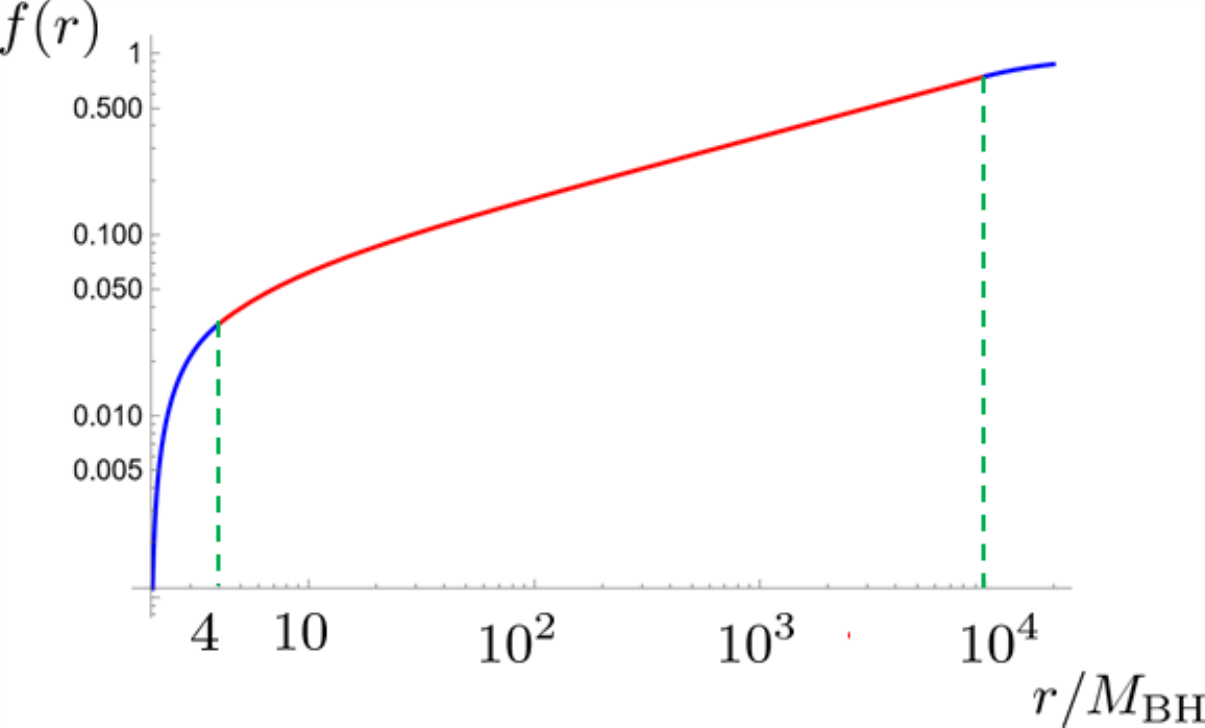}
\\[1em]
\noindent
\includegraphics[width=7cm]{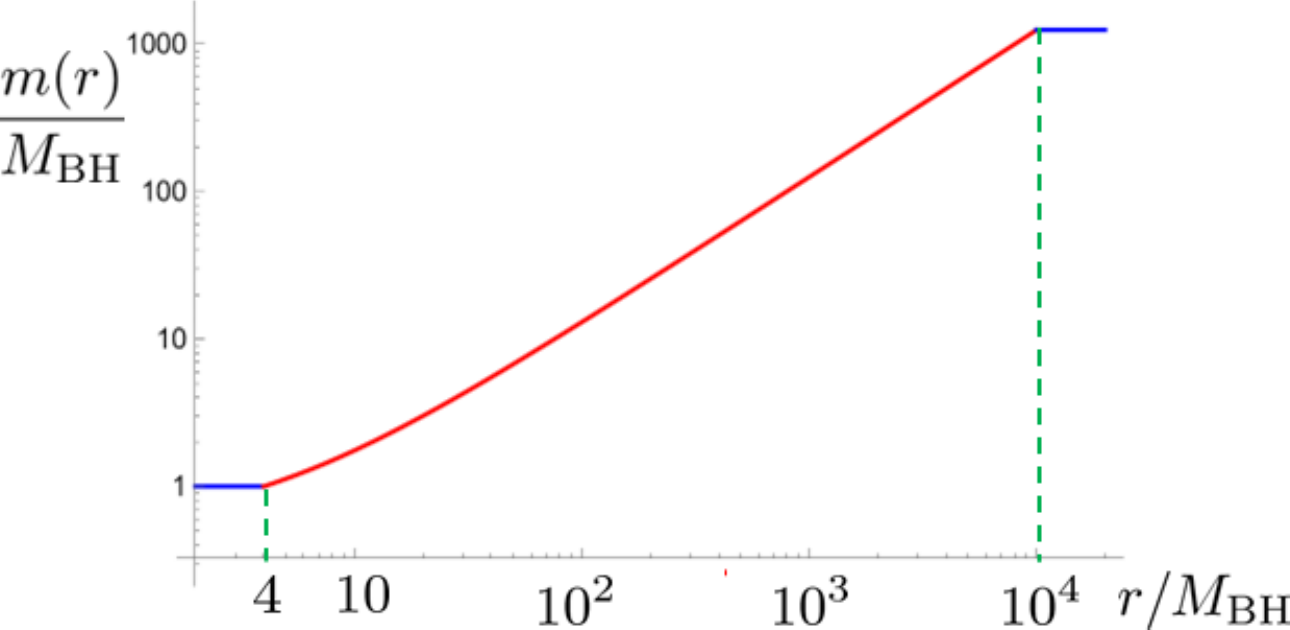}
\caption{Metric functions  for Model I\,, $f(r)$ (top)  and $m(r)$ (bottom).
We set $r_{\rm I}=4M_{\rm BH}\, (\alpha={1\over 8})$ and $r_{\rm O}=10^4 M_{\rm BH}$, which gives
$ M=1250.5 M_{\rm BH}$.}
\label{metric}
\end{figure}

\begin{figure}[ht]
\includegraphics[width=7cm]{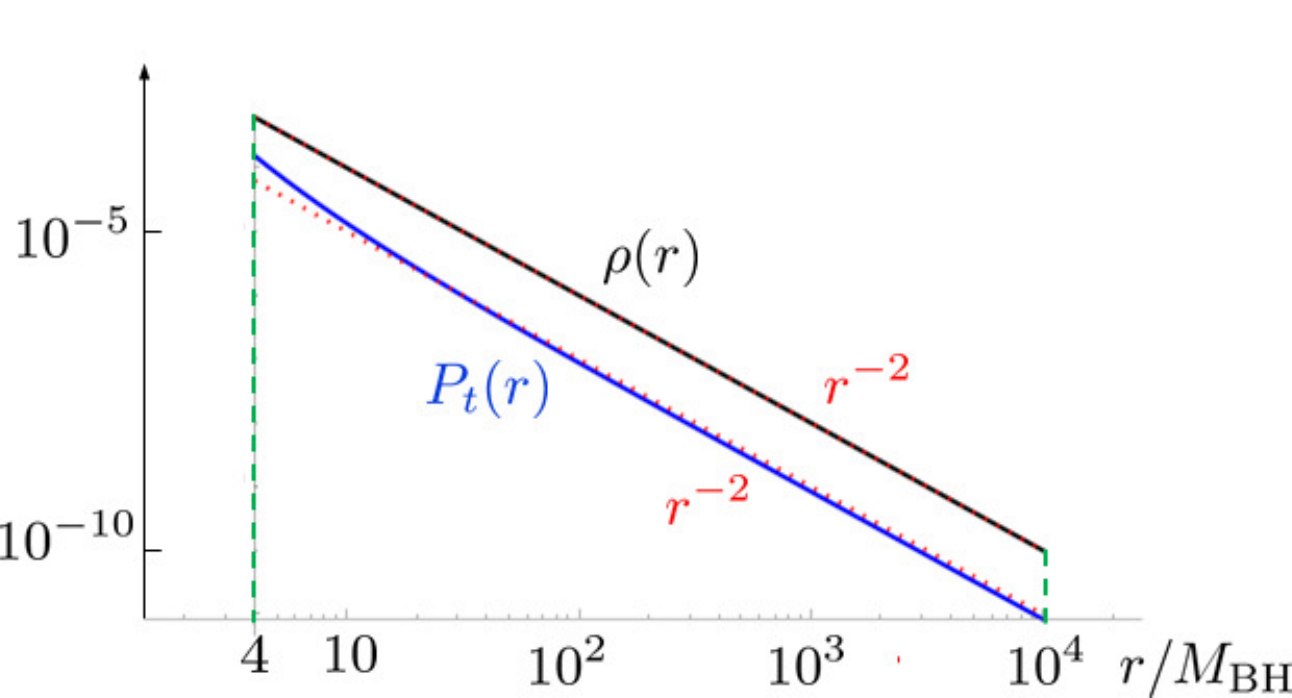}
\caption{The distributions of the energy density $\rho$ and tangential pressure $P_t$ for Model I. The parameters are the same as those in Fig. \ref{metric}.
}
\label{fig:ModelI}
\end{figure}

The metric functions $f(r)$ and $m(r)$ of the present model are shown in Fig.~\ref{metric}, and the energy density $\rho$ and tangential pressure $P_t$ in Fig.~\ref{fig:ModelI}.
We set $\alpha=1/8$, which gives $r_{\rm I}=4M_{\rm BH}$. We choose $r_{\rm O}=10^4M_{\rm BH}$. These parameters give $M=1250.5M_{\rm BH}$, $r_*={4\over 3}M_{\rm BH}$
 $f_0 \approx 0.0644$ and $c_0\approx 0.0348$.

\subsection{Model II}
The distribution above, with a very specific choice of ISCO radius, allows for marginally stable timelike circular geodesics in the entire spacetime. We also find a curious, analytic model that doesn't fine tune the integration constants, and for which every point in the matter distribution is an inflection point. We have to solve Eq. (\ref{eq_inflection}) for $r_{\rm I}\leq r\leq r_{\rm O}$.
Imposing $r>3m(r)$, we obtain 
\bea
m(r)={r\over 3[1+ (r/r_c)^2]}\,,
\ena
where 
$r_c$ is an integration constant. The density distribution is
\beann
\rho={m'\over 4\pi r^2}={1-(r/r_c)^2\over 12\pi r^2[1+ (r/r_c)^2]^2}\,.
\enann

The density vanishes at $r=r_c$, thus the outer radius of the distribution must be $r_{\rm O}\leq r_c$.
From the stability condition, i.e., 
 the outer radius is larger than the ISCO radius of the total mass $M_{\rm O}$ 
($r_{\rm O}\geq 6M_{\rm O}=6m(r_{\rm O})$), we find 
\beann
r_{\rm O}\geq {2r_{\rm O}\over [1+ (r_{\rm O}/r_c)^2]}
\,,
\enann
which gives $r_{\rm O}\geq r_c$.
As a result, the outer radius must be $r_{\rm O}=r_c$.

For the metric function $f$, we obtain
\beann
f={c_0 r^2\over 1+3 (r/r_{\rm O})^2}
\,,
\enann
with $c_0$ an integration constant.

Since the outside ($r>r_{\rm O}$) and inside ($r<r_{\rm I}$) 
of the Einstein cluster are assumed to be in vacuum, the continuity conditions at $r_{\rm I}$ and $r_{\rm O}$ yield,
\beann
m&=&\left\{
\begin{array}{cl}
1&(r<r_{\rm I})
\\[.5em]
{r\over 3[1+ (r/r_{\rm O})^2]}
&(r_{\rm I}<r<r_{\rm O})\,\,,
\\[.5em]
M&(r>r_{\rm O})
\\
\end{array}
\right.
\\
f&=&\left\{
\begin{array}{cl}
f_0
\left(1-{2\over r}\right)&(r<r_{\rm I})
\\[.5em]
{c_0 r^2\over (1+3(r/r_{\rm O})^2}&(r_{\rm I}<r<r_{\rm O})\,\,,
\\[.5em]
1-{2M\over r}&(r>r_{\rm O})
\\
\end{array}
\right.
\enann
where
\be
f_0={8\left(1+\beta^2 \right)\over  
\left(3+\beta^2 \right)^2 }\,,\quad
c_0={8\over 3r_0^2}\,.
\ee
Here we set $\beta\equiv r_{\rm O}/r_{\rm I}$.
The tangential pressure is given by
\beann
P_t={1-(r/r_{\rm O})^2\over
24\pi r^2(1+3(r/r_{\rm O})^2)(1+(r/r_{\rm O})^2)^2}\,.
\enann

The ISCO radius $r_{\rm I}$ is given by $\beta$, from the definition $m(r_{\rm I})=1$,
as
\beann
r_{\rm I}=3\left(1+ {1\over \beta^2}\right)
\,.
\enann
While the total mass $M$ is also described  by $\beta$ as
\beann
M=m(r_{\rm O})={r_{\rm O}\over 6}={\beta r_{\rm I}\over 6}={1+\beta^2 \over 2\beta }\,.
\enann
Since $\beta>1$, we find $3<r_{\rm I}<6$ and $1<M<\infty$.

\begin{figure}[ht]
\includegraphics[width=7cm]{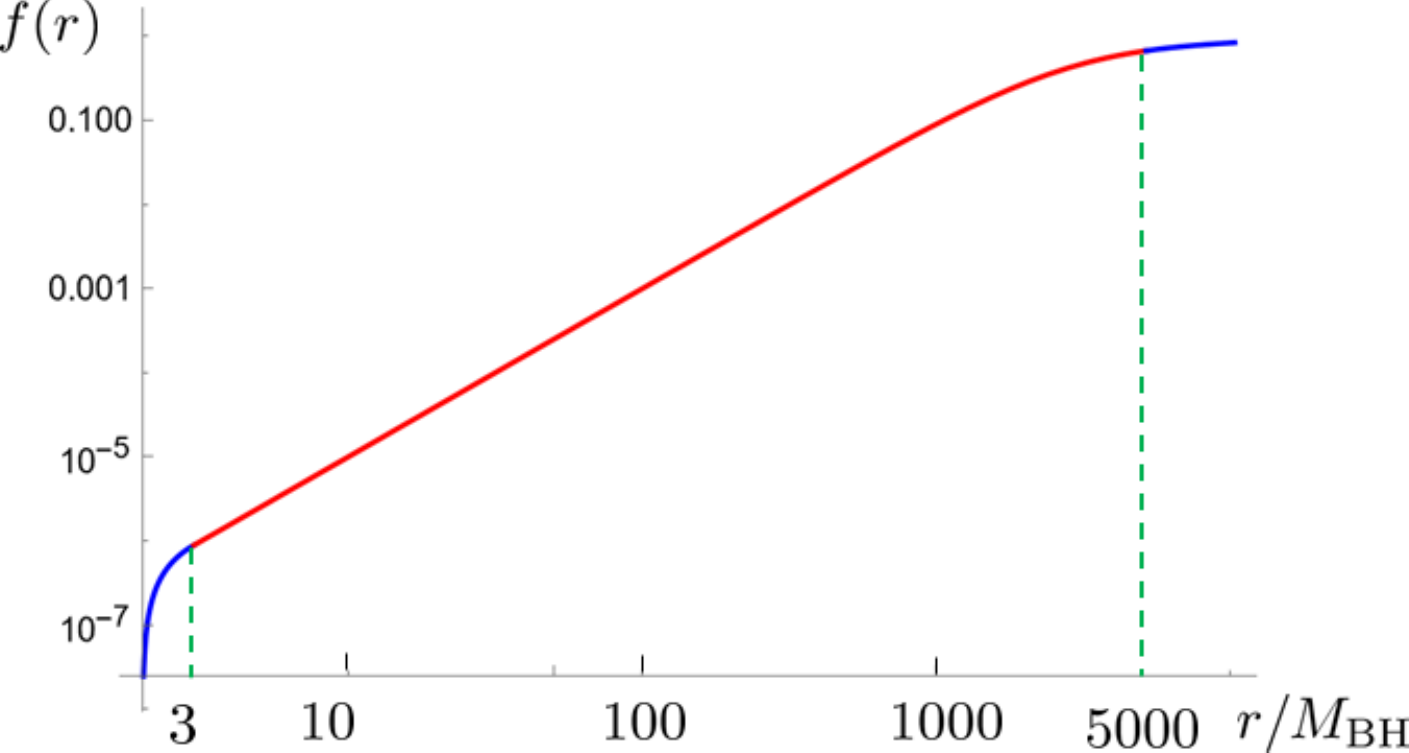}
\\[1em]
\noindent
\includegraphics[width=7cm]{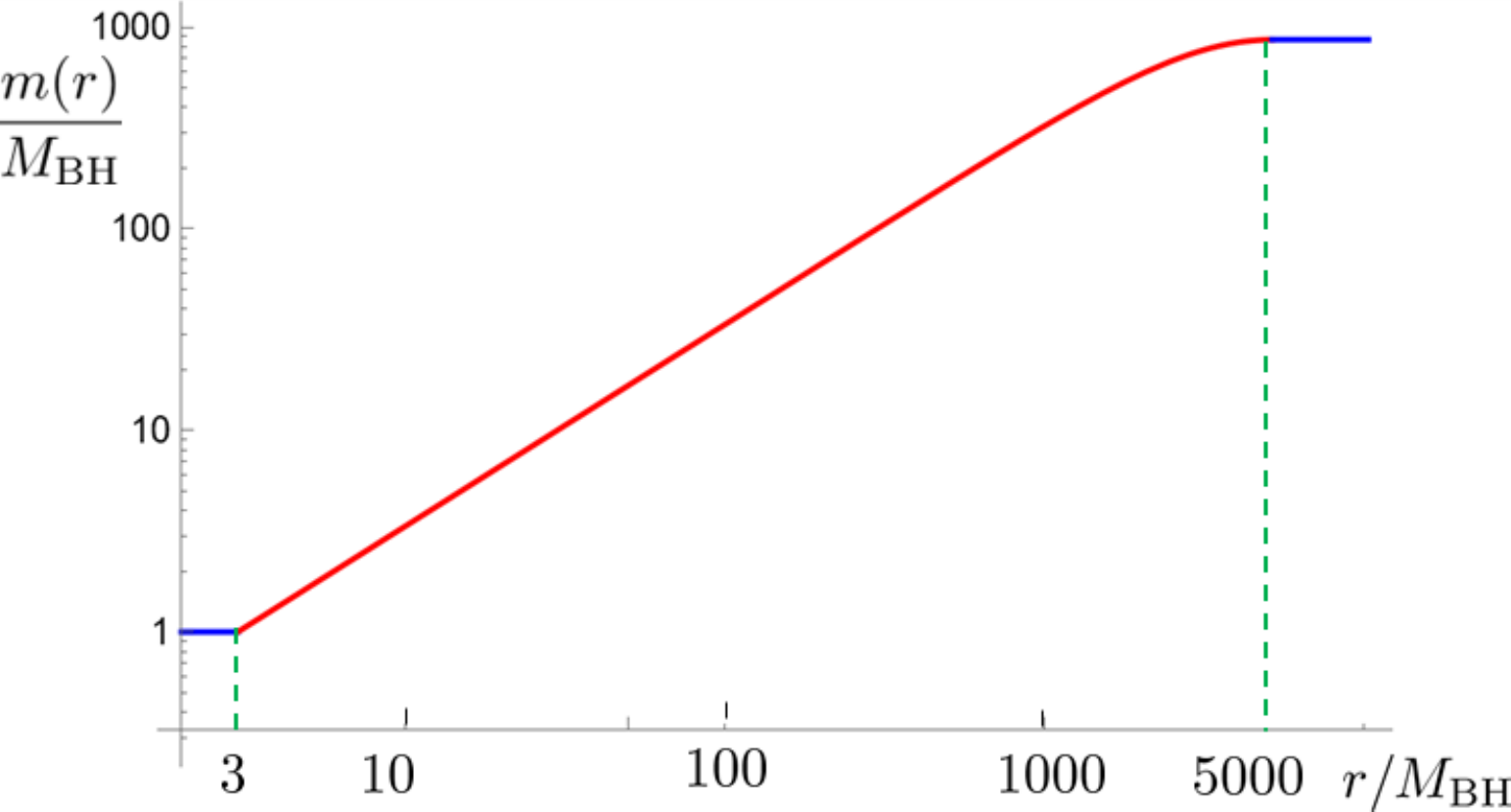}
\caption{Metric functions  for Model II\,, $f(r)$ (top)  and $m(r)$ (bottom).
We set $r_{\rm I}=3.000001M_{\rm BH}$, which gives  $r_{\rm O}=5196.15M_{\rm BH}$
and $M_{\rm O}=866.0 M_{\rm BH}$.}
\label{metric2}
\end{figure}
\begin{figure}[ht]
\includegraphics[width=7cm]{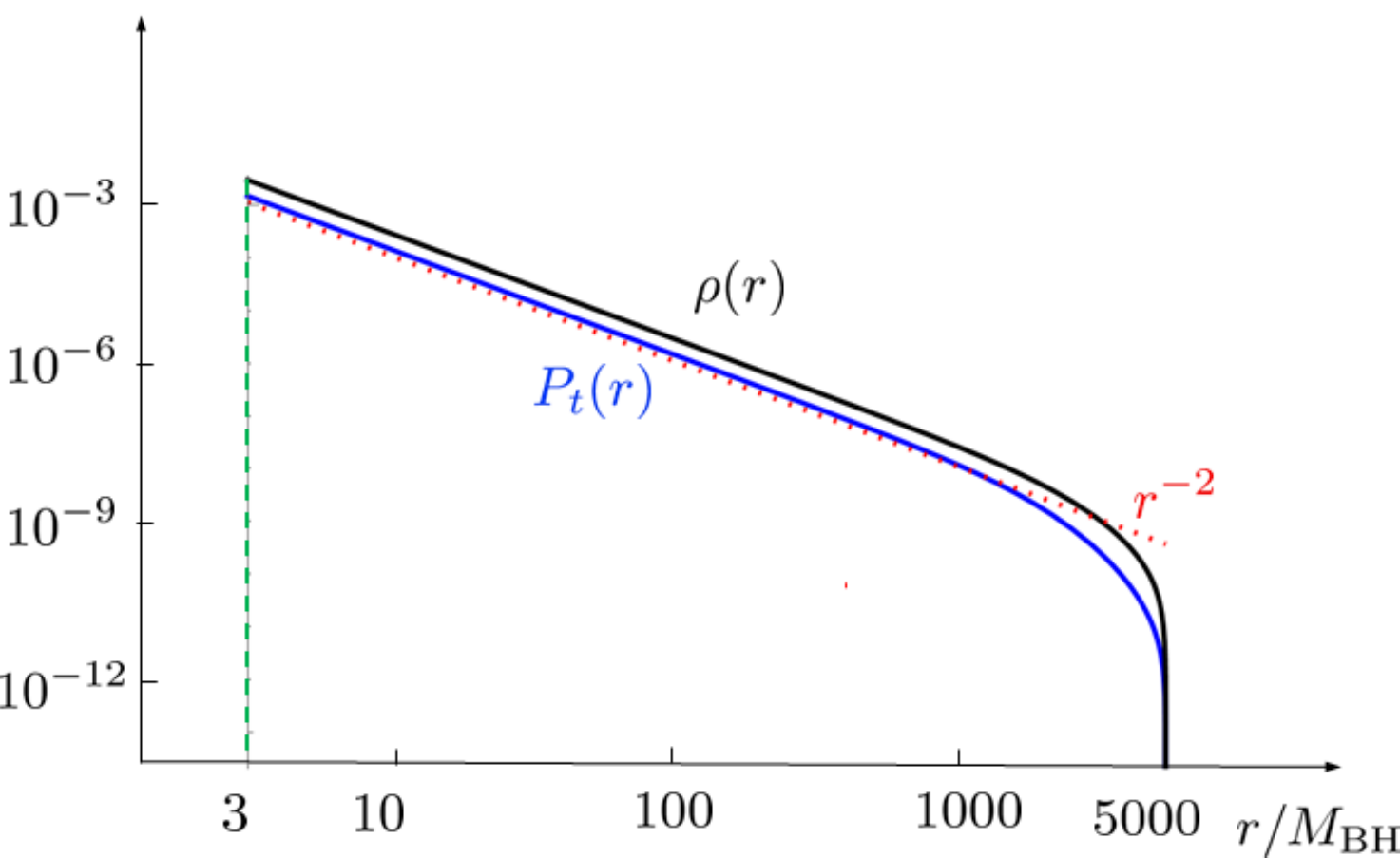}
\caption{The distributions of the energy density $\rho$ and tangential pressure $P_t$ for Model II. The parameters are the same as those in Fig. \ref{metric2}.
}
\label{rPt2}
\end{figure}
Expanding  the mass function $m(r) $ around $r_{\rm I}$, we find 
\beann
m(r)&\approx &1+\alpha(r-r_{\rm I})+O((r-r_{\rm I})^2)\,,
\enann
where
\beann
\alpha=
{r_{\rm O}^2(r_{\rm O}^2-r_{\rm I}^2)\over 3(r_{\rm O}^2+r_{\rm I}^2)^2}
={\beta^2(\beta^2-1)\over 3(\beta^2+1)^2}
\,,
\enann
which satisfies $0<\alpha<1/3$.
In the limit of $\beta\rightarrow \infty$, we find $\alpha\rightarrow 1/3$, i.e., 
$r_I\rightarrow 
3$ and $M\rightarrow \infty$.

We show the metric functions $f(r)$ and $m(r)$ of the present model  in Fig. \ref{metric2}, and 
the energy density $\rho$ and tangential pressure $P_t$ in Fig. \ref{rPt2}.
We set $\beta=3$, which gives $r_{\rm I}=10/3$, $r_{\rm O}=10$, and
 $M=5/2$.

\subsection{Model III}
Finally, we find another simple analytic model. We assume the mass function $m(r)$ as
\bea
m(r)=m_p r^p~~(r_{\rm I}\leq r\leq r_{\rm O})
\,,
\ena
where $m_p$ and $p(>0)$ are some constants.
The continuity of mass function at $r_{\rm I}$ is described by  $m(r_{\rm I})=m_p r_{\rm I}^p=1$, 
and the condition of the ISCO, 
\beann
r_{\rm I}^2 m'(r_{\rm I})+r_{rm I}m(r_{\rm I})-6m^2(r_{\rm I})=0\,,
\enann
gives 
\beann
(p+1)m_p r_{\rm I}^{p+1}-6m_p^2r_{\rm I}^{2p}=0
\,.
\enann
From these two conditions,
we find 
\beann
m_p=r_{\rm I}^{-p}\,,~{\rm and}~~p={6-r_{\rm I}\over r_{\rm I}}
\,.
\enann
The condition of 
$3<r_{\rm I}<6$, in which we are interested,  
gives $0<p<1$. The stability condition (\ref{stability}) gives 
\beann
r\geq \left({6m_p\over p+1}\right)^{1\over 1-p}=r_{\rm I}\,,
\enann
which is always satisfied.

The metric function should satisfy
\beann
{1\over f} {df \over d r}&=&{2m\over r(r-2m)}=
{2m_pr^{p-2}\over (1-2 m_p r^{p-1})}
\\
&=&{1\over 1-p} {d\over d r}\ln | (1-2 m_p r^{p-1})|
\,,
\enann
which is integrated as
\beann
f
=c_0 \left(1-{2m_p \over   r^{1-p}}\right)^{1\over 1-p}
\,,
\enann
where $c_0$ is an integration constant.

With the continuity conditions at $r=r_{\rm I}$ and $r_{\rm O}$, we find the metric functions as
\beann
m&=&\left\{
\begin{array}{cc}
1&(r<r_{\rm I})
\\
m_p r^p
&(r_{\rm I}<r<r_{\rm O})\,,
\\
M&(r>r_{\rm O})
\\
\end{array}
\right.
\\
f&=&\left\{
\begin{array}{cc}
f_0\left(1-{2\over r}\right)&(2M<r<r_{\rm I})
\\
c_0 \left(1-{2m_p \over   r^{1-p}}\right)^{1\over 1-p}
&(r_{\rm I}<r<r_{\rm O})\,,\\
1-{2M\over r}&(r>r_{\rm O})
\\
\end{array}
\right.
\enann
where
\beann
c_0&=&\left(1-{2M\over r_{\rm O}}\right)\left(1-{2m_p \over  r_{\rm O}^{1-p}}\right)^{-{1\over 1-p}}\,,
\\
f_0&=&{\left(1-{2M \over r_{\rm O}}\right)\over \left(1-{2\over r_{\rm I}}\right)}
\left[{\left(1-{2m_p \over  r_{\rm I}^{1-p}}\right)\over \left(1-{2m_p \over  r_{\rm O}^{1-p}}\right)}\right]^{{1\over 1-p}}\,.
\enann

In the present model, we have two 
 free parameters, $r_{\rm I}$ and $r_{\rm O}$, which should satisfy
\beann
&&
3< r_{\rm I}<6\,,
\\
&&
r_{\rm O}>r_{\rm I}
\,,
\enann
which gaurantees $M>1$.
$\alpha$ is given by
\beann
\alpha=m'(r_{\rm I})=p m_p r_{\rm I}^{p-1}
=p m_p^{1\over p}={6-r_{\rm I}\over r_{\rm I}^2}
\,,
\enann
which gives
\beann
0<\alpha <{1\over 3}
\,.
\enann

The energy density is given by
\beann
\rho={m'\over 4\pi r^2}={6-r_{\rm I}\over 4\pi r_{\rm I}^{p+1}}r^{3-p}\,.
\enann
The power exponent of density distribution ($3-p$) is uniform and its value takes between 2 and 3.

\begin{figure}[ht]
\includegraphics[width=7cm]{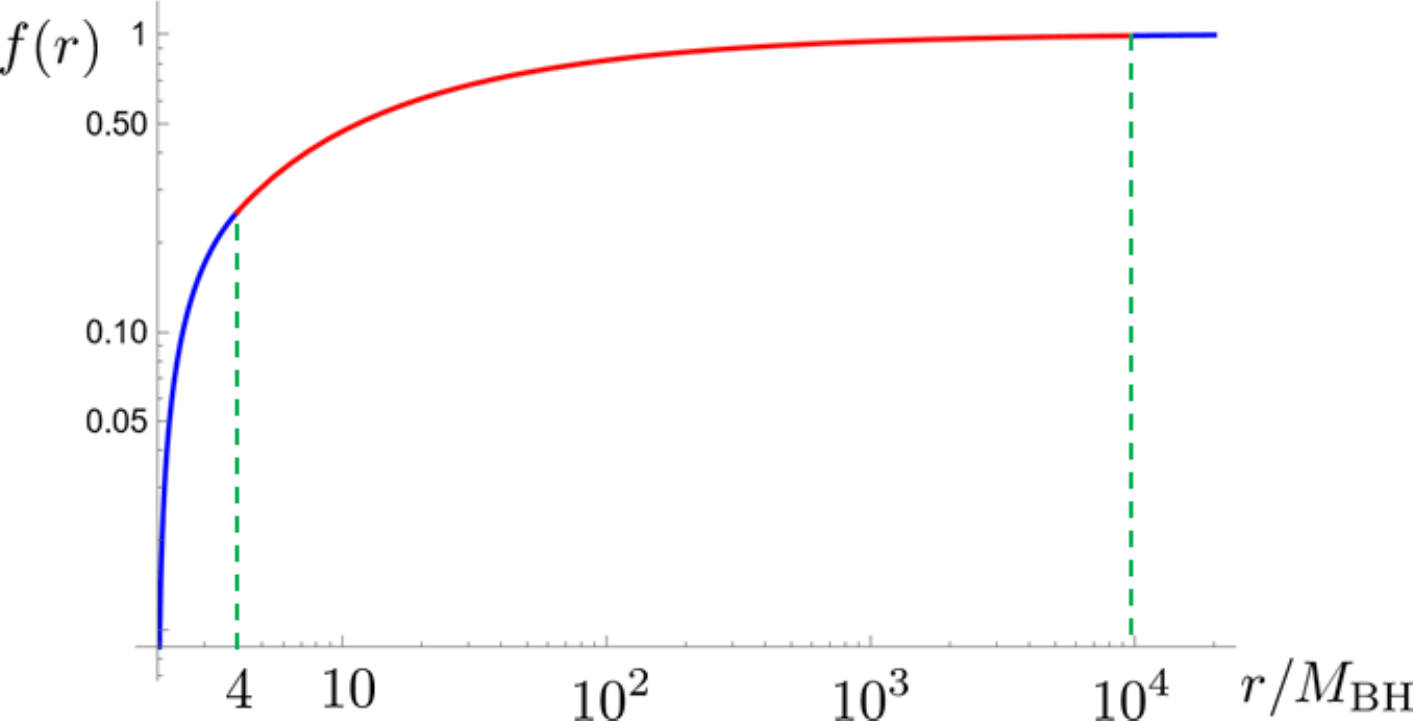}
\includegraphics[width=7cm]{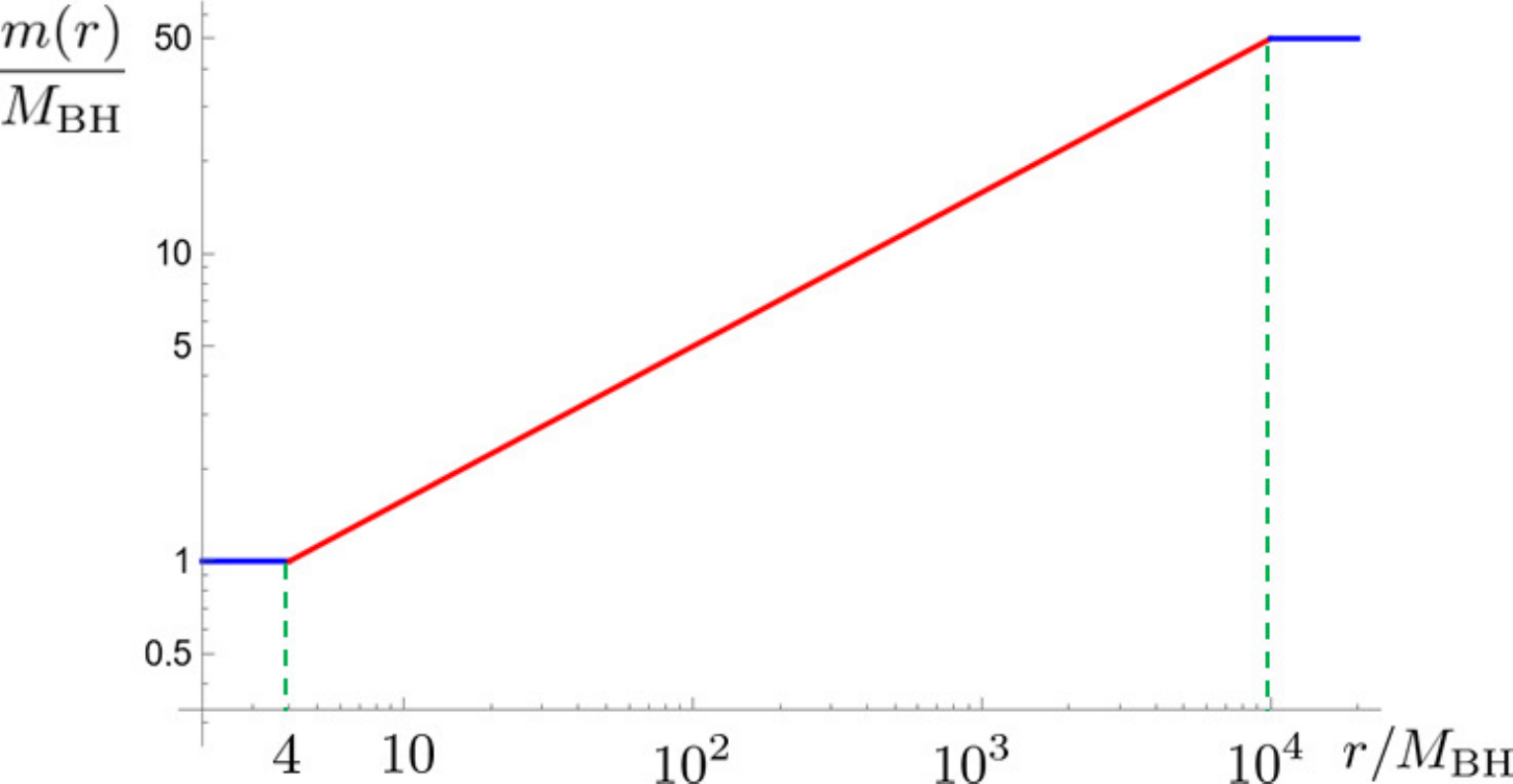}
\caption{Metric functions  for Model III\,, $f(r)$ (top)  and $m(r)$ (bottom).
We set $r_{\rm I}=4M_{\rm BH} (p=0.5)$ and $r_{\rm O}=10^4 M_{\rm BH}$, which gives
$ 50 M_{\rm BH}$. }
\label{metric3}
\end{figure}

\begin{figure}[ht]
\includegraphics[width=7cm]{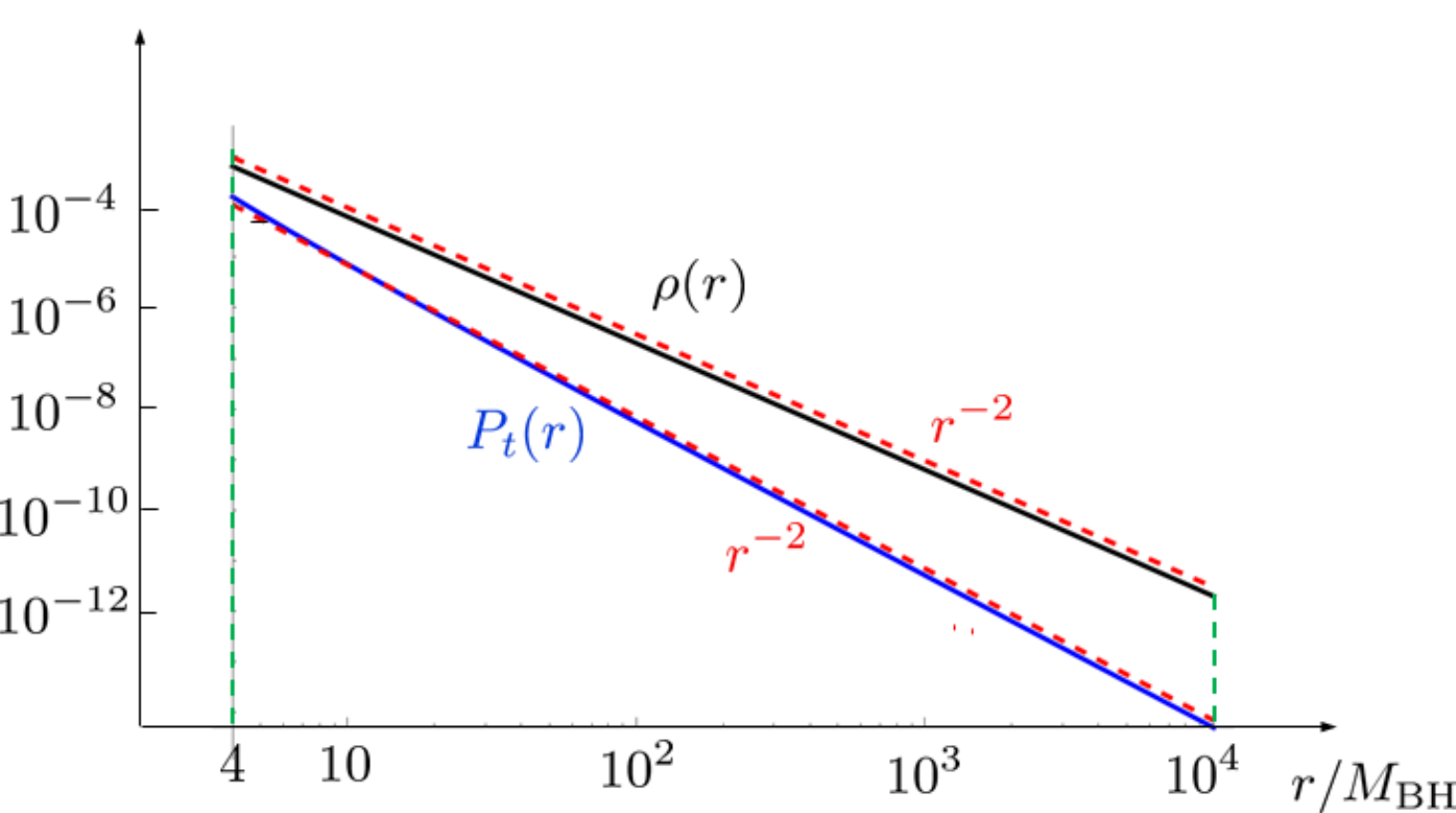}
\caption{The energy density $\rho$ and tangential pressure $P_t$ for Model III.
The parameters are the same as those in Fig.\ref{metric3}.}
\label{rhoPt3}
\end{figure}

We just show one example in Figs. \ref{metric3} and \ref{rhoPt3}.

This model can be extended to the similar mass function such that
\bea
m={m_pr^p\over 1+(r_*/r)^{(1-p)}}\,,
\ena
with $0<p<1$.
The metric function is 
\bea
f=c_0\left[1-{2m_p-r_*^{1-p}\over r^{1-p}}\right]^q
\,,
\ena
where 
\beann
q={1\over (1-p)\left[1-r_*^{1-p}/(2m_p)\right]}
\,.
\enann

For $r_*=0$, we recover  the previous model.
The energy density is given by
\bea
\rho={m'\over 4\pi r^2}
={pm_p\over 4\pi r^{3-p}}{1+{2-p\over p}\left({r_*\over r}\right)^{1-p}\over 
\left[1+\left({r_*\over r}\right)^{1-p}\right]^2}\,.
\ena
For $r>r_*$, the density distribution is almost the same as the previous model, i.e., $\gamma=3-p$. In the range of $r<r_*$, the power exponent of the density distribution becomes $\gamma=2$.

\begin{widetext}

\subsection{Comparison of the three models}
We summarize the three toy models in Table \ref{Table:summary1}.
For Models I and III, $m(r)\rightarrow \infty $ as $r\rightarrow \infty$. Hence we have to terminate the distribution at a finite radius
$r_{\rm O}$.
Here we set $r_{\rm O}=10^4 M_{\rm BH}$.
As examples, the ISCO radii are chosen as  $r_{\rm I}=5.5 M_{\rm BH}\,, 4M_{\rm BH}$ and $3M_{\rm BH}$
(or a value close to it if not possible).

\begin{table*}[ht]
\begin{center}
  \begin{tabular}{|c||c|c||c|c||rl|c|}
\hline 
Model &$r_{\rm I}/M_{\rm BH} $&$r_{\rm O}/M_{\rm BH} $&$M/M_{\rm BH} $ &$M/r_{\rm O}$&$\gamma$~~&&properties
\\
\hline
\hline
& 5.5& &$166.2$ &$0.0166$ &  &&isothermal 
\\
\cline{2-2}\cline{4-5}
I&4& $10000$&$1250.5$ &0.125 &$2$~   &&
stable
\\
\cline{2-2}\cline{4-5}
& 3& &$3333.3$ &0.333 &   & & (marginally stable)
\\
\hline
\hline
&5.5 &6.025 &1.00416 &0.1667&~12~($\rightarrow$&$\infty$)~&
\\
\cline{2-7}
II&4 &6.928 &1.1547 &0.1667&~4~($\rightarrow$&$\infty$)~&
marginally stable
\\
\cline{2-7}
&3.001 &164.4 &27.4 &0.1667&~2~($\rightarrow$&$\infty$)~&too small $r_{\rm O}$ except for $r_{\rm I}\approx 3$
\\
\cline{2-7}
&3.000001 &5196.15 &866.0 &0.1667&~2~($\rightarrow$&$\infty$)&
\\
\hline
\hline
& & & & &&&\\[-1em]
&5.5\,($p={1\over 11}$)&& 42.6&0.00426&&&stable
\\[.1em]
\cline{2-2}\cline{4-5}
&& &  & &&&\\[-1em]
III&4\,($p={1\over 2}$)&10000 & 50&0.005&2.5~&&too small mass $M$
\\[.1em]
\cline{2-2}\cline{4-5}
&3\,($p=1$) & & 57.7&0.00577&&& 
\\
\hline
\hline
 \end{tabular}
\caption{Typical values of parameters in three toy models (I, II, and III). 
For Model I and III, we choose $r_{\rm O}=10^4 M_{\rm BH}$, while the radius $r_{\rm O}$
 of Model II is fixed.
The power exponent of 
the density distribution is defined by $\gamma=-{d\ln \rho\over d\ln r}$.
In Model II, the density drops rapidly near the outer boundary $r_{\rm O}$.}
\label{Table:summary1}
\end{center}
\end{table*}

\end{widetext}
\section{Some Realistic Model}
\label{realistic_model}
Although we do not know the exact DM distribution for any galaxy, a general feature is the existence of some characteristic galactic scale, where the power exponent $\gamma$ of the distribution changes~\cite{sand2004dark,Borriello:2000rv,Salucci:2018hqu}. This scale  is much larger than the SMBH horizon scale. Hence we consider some model with a typical scale length much larger than the horizon.

To model realistic DM distributions more closely, we consider the following model with mass function
\bea
m(r)={m_0+m_1 r+M r^2\over (r+r_*)^2}\,,
\label{mass function IV}
\ena
where $m_0, m_1,$ and $M$ are some constants. We call it Model IV.

Inside $r_{\rm I}$ is a vacuum and there exists a BH with the mass $M_{\rm BH}=1$. From the continuity of the mass function $m(r)$, we find $m(r_{\rm I})=1$. The inner boundary $r_{\rm I}$ is assumed to be the ISCO radius, which satisfies
\beann
r_{\rm I}^2 m'(r_{\rm I})+r_{\rm I} m(r_{\rm I})-6m^2(r_{\rm I})=0
\,,
\enann
with
\bea
m'(r)={m_1 r_*-2m_0+(2Mr_*-m_1)r\over (r+r_*)^3}\,.
\ena

The above two conditions give
\bea
&&
m_0+m_1 r_{\rm I}+M r_{\rm I}^2=(r_{\rm I}+r_*)^2 
\label{condition1}
\\
&&
r_{\rm I}^2\left[m_1 r_*-2m_0+(2Mr_*-m_1)r_{\rm I} \right]
\nn
&&~~
=(6-r_{\rm I})(r_{\rm I}+r_*)^3\,,
\label{condition2}
\ena
which solve $m_0$ and $m_1$ as
\bea
m_0&=&Mr_{\rm I}^2+{2(r_*+r_{\rm I})\over r_{\rm I}}\left[(r_1-3)r_*-3r_{\rm I}\right]\,,
\label{m0}\\
m_1&=&{(r_*+r_{\rm I})\over r_{\rm I}^2}\left[(6-r_1)r_*+(6+r_{\rm I})r_{\rm I}\right]-2Mr_{\rm I}\,.
\nn
&&
\label{m1}
\ena

\begin{widetext}

For Einstein cluster, since $L^2>0$ and $E^2>0$, we have the  constraint of
$r>3m(r)$. 
The stability condition (\ref{stability}) should be also satisfied. 
Hence we have two constraints, 
 which are desribed as
 \bea
 C_1(r) &\equiv &r^3+(2r_*-3M)r^2+(r_*^2-3m_1)r-3m_0>0\,,
 \label{constraint1}
 \\
 C_2(r)   &\equiv&
Mr^5+2M(2r_*-3M)r^4+(3Mr_*^2+2m_1r_*-12m_1M-m_0)r^3
 \nn
 &&~~-2(6m_0M+3m_1^2-m_1r_*^2)r^2+m_0(r_*^2-12m_1)r-6m_0^2\geq 0\,.
 \label{constraint2}
\ena
 These constraints should be satisfied in the whole range of distribution ($r_{\rm I}\leq r\leq r_{\rm O}$).
 This restricts the parameter range of $r_*$ and $r_{\rm I}$ (and $M$ if it is not fixed).
Note that those constraints are satisfied at $r_{\rm I}$ as
\beann
C_1(r_{\rm I})=(r_{\rm I}-3)(r_{\rm I}+r_*)^2 >0\,,~~C_2(r_{\rm I})=0
\,,
\enann
because  $r_{\rm I}>3$. Since $C_2(r)$ vanishes at $r_{\rm I}$, we can factorise it with $(r-r_{\rm I})$ by use of 
Eqs. (\ref{m0}) and  (\ref{m1}). 
Hence setting $
{C}_2(r)=(r-r_{\rm I})
\tilde{C}_2(r)
$, 
the constraint with $C_2(r)$ is reduced to a quartic inequality\, i.e., 
\bea
\tilde{C}_2(r)\geq 0
\,,
 \label{constraint22}
\ena
where 
\bea
\tilde{C}_2(r)&\equiv& Mr^4-M(6M-4r_*-r_{\rm I})r^3
\nn
&&
+{1\over r_{\rm I}^2}
\left[6(r_*+r_{\rm I})^2(2r_*+r_{\rm I})-2r_*^2r_{\rm I}(r_*+r_{\rm I})+18M^2r_{\rm I}^3
-3M(24(r_*+r_{\rm I})^2-4r_{\rm I}(r_*^2-r_{\rm I}^2)-r_*^2r_{\rm I}^2)\right]r^2
\nn
&&
-{1\over r_{\rm I}^4}\left[6(r_*+r_{\rm I})^2-2r_*r_{\rm I}(r_*+r_{\rm I})-Mr_{\rm I}^3\right]\left[
36(r_*+r_{\rm I})^2+12r_{\rm I}^2(r_*+r_{\rm I})-r_*^2r_{\rm I}^2-18Mr_{\rm I}^3\right]r
\nn
&&+{6\over r_{\rm I}^3}\left[6(r_*+r_{\rm I})^2-2r_*r_{\rm I}(r_*+r_{\rm I})-Mr_{\rm I}^3\right]^2
\,.
\ena

 \end{widetext}

 The energy density $\rho$ is 
 \beann
 \rho={m'\over 4\pi r^2}
 ={m_1 r_*-2m_0+(2Mr_*-m_1)r\over 4\pi r^2 (r+r_*)^3}
 \,.
 \enann
 From the condition (\ref{condition2}), we find $r_{\rm I}\leq 6$ because $\rho(r_{\rm I})\geq 0$.
 If $\rho(r_{\rm I})=0$, we find $r_{\rm I}=6$, which is the BH ISCO radius.
 Since we are interested in the case of $r_{\rm I}<6$, 
 the energy density does not vanish at $r_{\rm I}$.

 The metric function $f$ is obtained as
 \bea
 f=f_0\exp I(r)\,,
 \ena
 where $f_0=f(r_{\rm I})$ is an integration constant and 
 $I(r)$ is defined by the integration 
 \beann
 I(r)&\equiv& \int_{r_{\rm I}}^r dr {2m(r)\over r(r-2m(r))}
\\
& =&
\int_{r_{\rm I}}^r dr {2(m_0+m_1 r+M r^2)\over r\left[r(r+r_*)^2-2(m_0+m_1 r+M r^2)\right]}
 \,,
 \enann
which analytic solutions are given  in Appendix \ref{Appendix}.
For the models we consider here, the metric function $f(r)$  is always finite in the whole range.

 From the density distribution, 
 we can classify this model into two  cases.
 \begin{itemize}
  \item  Type A : $m_1 r_*-2m_0=0$\\
  Since the density must be positive, $(2Mr_*-m_1)$ is positive.
  \item Type B : $m_1 r_*-2m_0\neq 0$ \\
  This typel is further classified into three cases as\\
\begin{itemize}
  \item[*] Type B$_+$ : $2Mr_*-m_1> 0$
  \item[*] Type B$_0$ : $ 2Mr_*-m_1=0$
  \item[*] Type B$_-$ : $2Mr_*-m_1< 0$
\end{itemize}
In the last two cases, $(m_1 r_*-2m_0)$ must be positive.
Since the density vanishes at a finite radius in Type B$_-$, 
the radius of the distribution  must be finite.
\end{itemize}

We shall discuss these types separately in below.

\subsection{\underline{\bf Type A}}
In the  case of $m_1 r_*-2m_0=0$, there are two free parameters $r_*$ and $r_{\rm I}$.
The other parameters $m_0, m_1, M$ are fixed as
\bea
m_0
&=&{r_*\over 2r_{\rm I}}\left[3(r_{\rm I}-2)r_*-(6-r_{\rm I})r_{\rm I}\right]
\label{Hm0}
\,,
 \\
m_1
&=&{1\over r_{\rm I}}\left[3(r_{\rm I}-2)r_*-(6-r_{\rm I})r_{\rm I}\right]
\label{Hm1}
\,,
\\
 M
&=&1+{(6-r_{\rm I})\over 2r_{\rm I}^3}(r_*+r_{\rm I})(r_*+2r_{\rm I})
\,.
\label{HM}
\ena

The density distribution is 
\beann
 \rho={2Mr_*-m_1\over 4\pi r (r+r_*)^3}
 \,,
 \enann
which is the Hernquist type distribution, i.e., $\rho \propto r^{-\gamma}$ 
with $\gamma=1$ for $r<r_*$ while $\gamma=4$ for $r>r_*$.

Since $M>1(=M_{\rm BH})$, $r_{\rm I}<6$.
The condition of $2Mr_*-m_1>0$ gives
\beann
C_3\equiv (6-r_{\rm I})(r_*+r_{\rm I})^3-2r_*r_{\rm I}^3>0
\,.
\enann

\begin{figure}[ht]
\includegraphics[width=8cm]{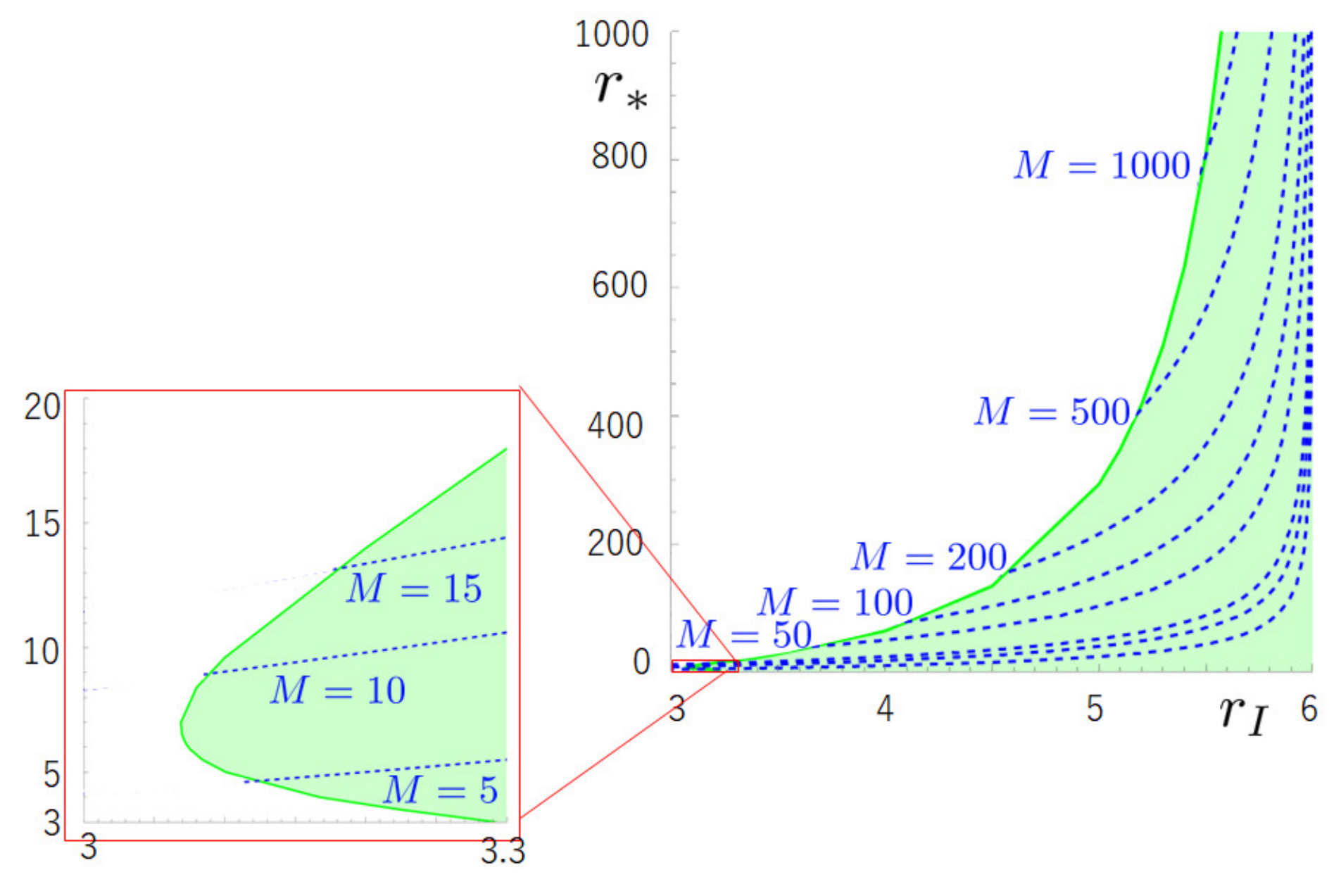}
\caption{The existence range of solutions of Type A in the $r_{\rm I}$-$r_*$ plane. 
The left panel is the enlarged one of the left-bottom corner.
The blue dotted curves are contours of mass $M$.}
\label{rIr0A}
\end{figure}

As  we discussed, we have two more constraints on $r_*$ and $r_{\rm I}$
by  $C_1(r)>0$ and $\tilde C_2(r)>0$ 
since $M$ is given by Eq. (\ref{HM}).
These three constraints restrict 
the parameter range of $r_*$ and $r_I$.
We show it 
by the lightgreen region in Fig. \ref{rIr0A}.
For larger values of $r_*$, the existence region in $r_{\rm I}$ becomes 
smaller and $r_{\rm I}\rightarrow 6$ as $r_*\rightarrow \infty$.

Since M is the total mass of a galaxy, 
the scale of the gravitational potential  is $M/r*$, which is 
given by 
\beann
{M\over r_*}&=&{1\over r_*}\left[1+{(6-r_{\rm I})\over 2r_{\rm I}^3}(r_*+r_{\rm I})(r_*+2r_{\rm I})
\right]\,.
\enann
It is approximated as 
\beann
{M\over r_*}
&\sim& {(6-r_{\rm I})\over 2r_{\rm I}^3}
\,,
\enann
when $r_*\gg 1$.

\subsection{\underline{\bf Type B$_0$}}
In the case of $2Mr_*-m_1=0$, there are also two free parameters $r_*$ and $r_{\rm I}$.
The other parameters $m_0, m_1, M$ are fixed as
\bea
m_0
&=&{2r_*^2(r_{\rm I}-3)\over r_{\rm I}}-{1\over 2}(6-r_{\rm I})(3r_*+r_{\rm I})
\,,~~~~~~
 \\
m_1
&=&{r_*\over r_{\rm I}^2}\left[(6-r_I)r_*+(6+r_{\rm I})r_{\rm I}\right]
\,,
\\
 M
&=&{1\over 2r_{\rm I}^2}\left[(6-r_I)r_*+(6+r_{\rm I})r_{\rm I}\right]
\,.
\ena

\begin{figure}[ht]
\includegraphics[width=7cm]{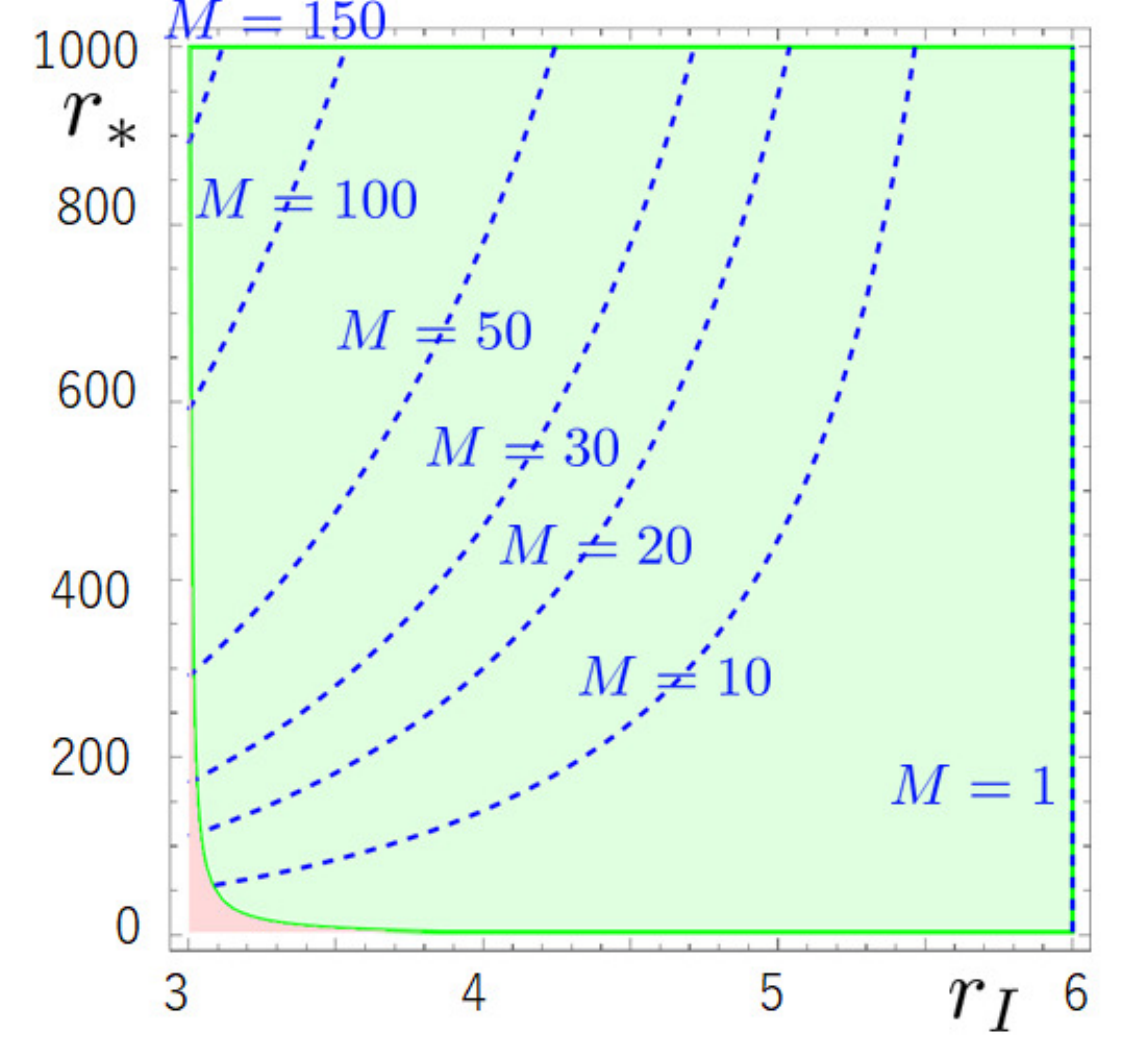}
\caption{The existence range Type B$_0$ of solutions in the $r_{\rm I}$-$r_*$ plane. 
The blue dotted curves are contours of mass $M$.}
\label{rIr0B0}
\end{figure}

Since 
\bea
m_1 r_*-2m_0={(6 -r_{\rm I}) (r_* + r_{\rm I})^3\over r_{\rm I}^2} > 0 
\,,
\label{constraint4}
\ena
when $r_{\rm I}<6$, 
we have two constraints on $r_*$ and $r_{\rm I}$
by  $C_1(r)>0$ and $\tilde C_2(r)>0$.
These constraints restrict 
the parameter range of $r_*$ and $r_I$.
We show it 
by the lightgreen region in Fig. \ref{rIr0B0}.

~~
\\
\vskip 1cm

The energy density is given by
 \beann
 \rho={m'\over 4\pi r^2}
 ={m_1 r_*-2m_0\over 4\pi r^2 (r+r_*)^3}
 \,,
 \enann
i.e., $\rho \propto r^{-\gamma}$ 
with $\gamma=2$ for $r<r_*$ while $\gamma=5$ for $r>r_*$.

\subsection{ \underline{\bf Type B$_\pm$ }}
In Type  B$_\pm$, three parameters ($r_*\,, r_{\rm I}$\,, and $M$) are free.
The other parameters ($m_0\,, m_1$) are fixed by (\ref{m0}) and  (\ref{m1}).
The condition of $2Mr_*-m_1\gtrless 0$  gives the constraint on $M$
as 
\beann
M \gtrless  {1\over 2r_{\rm I}^2}\left[(6-r_{\rm I})r_*+(6+r_{\rm I})r_{\rm I}\right]\,.
\enann

With this constraint, two additional constraints (\ref{constraint1}) and (\ref{constraint22}) give 
the constraint on $M$ for given values of $r_{\rm I}$ and $r_*$.
In Fig. \ref{rIMB}, we show the existence region of solutions for
 given values of $r_*=10^2, 10^3$ and $10^4$.
The solutions of Type B$_+$ exist in  the lightgreen region in Fig. \ref{rIMB}.
The energy density extends to infinity.
The density profile is $\rho\propto r^{-\gamma}$ with 
$\gamma=2$ for $r<r_*$ while $\gamma=4$ for $r>r_*$.


While, for Type  B$_-$, the solutions exist in the green region in Fig. \ref{rIMB}.
The energy density vanishes at finite radius $r_{\rm O}$, which 
is

\beann
r_{\rm O}&=&-{m_1r_*-2m_0\over 2Mr_*-m_1}
\nn
&=&{\left[(6-r_{\rm I})r_*^2+3(6-r_{\rm I})r_{\rm I}r_*+12r_{\rm I}^2\right]-2Mr_{\rm I}^3
\over \left[(6-r_{\rm I})r_*+(6+r_{\rm I})r_{\rm I}-2Mr_{\rm I}^2\right]}
\,.
\enann
It should be 
the outer boundary of DM distribution.
The total mass of a galaxy is
\bea
m(r_{\rm O})={m_0+m_1r_{\rm O}+Mr_{\rm O}^2\over (r_{\rm O}+r_*)^2}=M_{\rm G}\,.
\ena

The density profile is $\rho\propto r^{-\gamma}$ with 
$\gamma=2$ for $r<r_*$ while $\gamma=5$ for $r>r_*$.
However the outer boundary $r_{\rm O}$ is almost same as 
$r_*$ when $r_*\gg 1$. As a result $\gamma$ is indefinite near the boundary.

Note that the solution of Type B$_0$ exists at the boundary between two types, B$_+$ and B$_-$.
The solution of Type A is also a special limit in Type  B$_+$, which is shown by the red curve.

\begin{widetext}

\begin{figure}[ht]
\includegraphics[width=15cm]{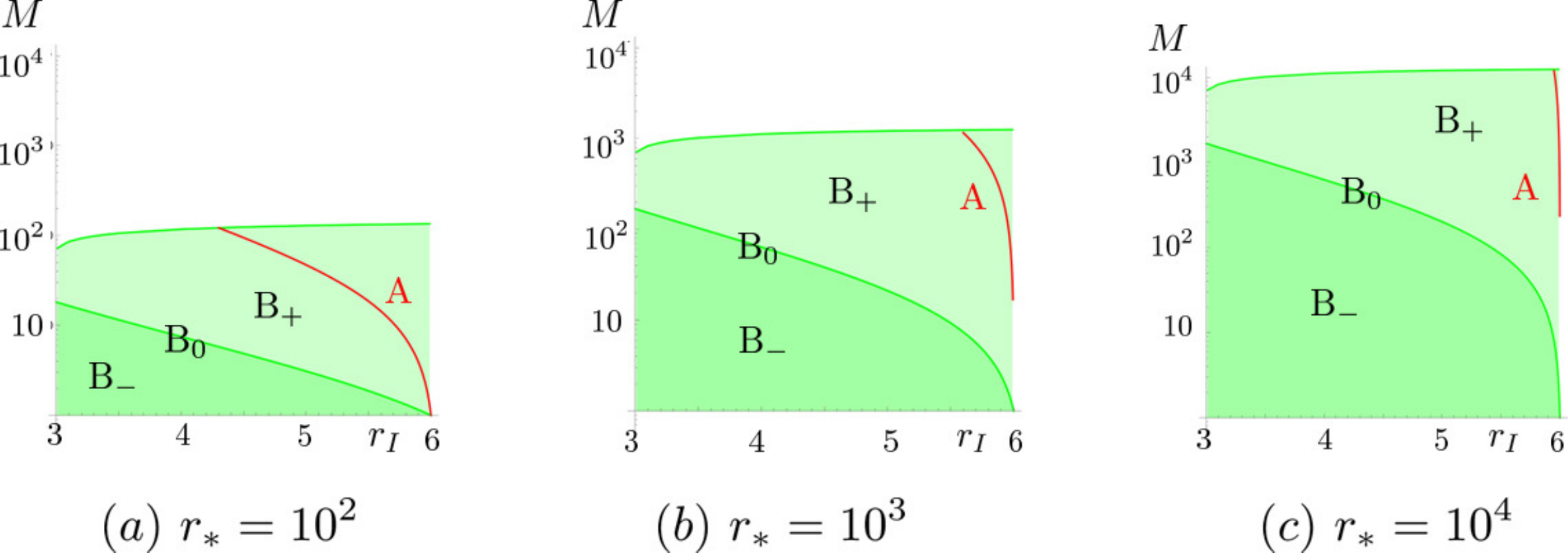}
\caption{The existence range of solutions in the $r_{\rm I}$-$M$ plane for Model IV.
The solutions of Type B$_+$ and Type B$_-$  exist in the lightgreen and green regions, respectively. We choose (a) $r_*=10^2$\,, (b) $r_*=10^3\,,$ and (c) $r_*=10^4$. The solutions of Type B$_0$ are found  at the boundary. The solutions of Type A (the red curve) appear  as a special limit of Type B$_+$.
}
\label{rIMB}
\end{figure}

\subsection{Typical values of Model IV and density distributions}
In Table \ref{Table:summary2}, we summarize the typical values of Model IV. 
\begin{table}[ht]
\begin{center}
\scalebox{0.8}[0.9]{
  \begin{tabular}{|c||c|c|c||c|c|c||rl|c|}
\hline 
Type &$r_{\rm I}/M_{\rm BH} $&$r_*/M_{\rm BH} $&$r_{\rm O}/M_{\rm BH} $&$M/M_{\rm BH} $ &$M/r_*$ &$M/r_{\rm O}$&$\gamma$&&DM distribution
\\
\hline
\hline
&5.5& 808&&1002.1 &1.240&&&&
\\
\cline{2-3}\cline{5-6}
A&4& 64&$\infty$&77.5 &1.211&0&1~$\rightarrow$&4&Hernquist profile
\\
\cline{2-3}\cline{5-6}
&3.069&6.9 &&7.59 &1.100&&&&
\\
\hline
\hline
&5.5& &&83.69&0.008369&&&&
\\
\cline{2-2}\cline{5-6}
B$_0$&4&10000&$\infty$&626.25 &0.062625&$0$&2~$\rightarrow$&5&infinite radius
\\
\cline{2-2}\cline{5-6}
&3& &&1668.17 &0.166817&&&&
\\
\hline
\hline
&5.5&& & 90&0.009&&&&
\\
\cline{2-2}\cline{5-6}
B$_+$&4&10000 &$\infty$ &700 &0.07&0&2~$\rightarrow$&4&infinite radius
\\
\cline{2-2}\cline{5-6}
&3&&& 1800&0.18&&&&
\\
\hline
\hline
&5.5& &$1.2\times 10^6$ & 83&0.0083&$6.92\times 10^{-5}$&&&
\\
\cline{2-2}\cline{4-7}
B$_-$&4&10000 &$2.5\times 10^7$ & 626&0.0626&$2.5\times 10^{-5}$&2~$\rightarrow$&5&
finite radius
\\
\cline{2-2}\cline{4-7}
&3&&$1.0\times 10^8$ & 1668&0.1668&$1.67\times 10^{-5}$&&&
\\
\hline
\hline 
 \end{tabular}
 }
\caption{Typical values of parameters in Model IV. 
The scale length of DM distribution is chosen as $r_*= 10^4 M_{\rm BH}$. }
\label{Table:summary2}
\end{center}
\end{table}
We choose $r_{\rm I}=5.5 M_{\rm BH}\,, 4M_{\rm BH}$ and
 $3M_{\rm BH}$
(or a value close to it if not possible).
In Type A, since $r_*$ is bounded above, we choose the maximum values for given $r_{\rm I}$. We also choose $r_{\rm I}=3.069M_{\rm BH}$, which is the minimum possible value.
For Type B, the scale length of DM distribution is chosen as $r_*= 10^4 M_{\rm BH}$.
For the masses of Type B$+$ and Type B$_-$, 
we choose a slightly larger and smaller values than 
the mass of Type B$_0$, respectively.


%

In Figs. \ref{densityAB}, we show the density distributions
 for these models in Table \ref{Table:summary2}. We choose $r_{\rm I}=4M_{\rm BH}$. 
The power exponent 
$\gamma$ changes from 1 to 4 for Type A, from 2 to 5 for Type B$_0$, from 2 to 4 for Type B$_+$, and from 2 to 5 for Type B$_-$, respectively. 
For Type B$_-$, the power exponent $\gamma$ diverges near the outer boundary because the density vanishes rapidly there.

\begin{figure}[ht]
\includegraphics[width=7cm]{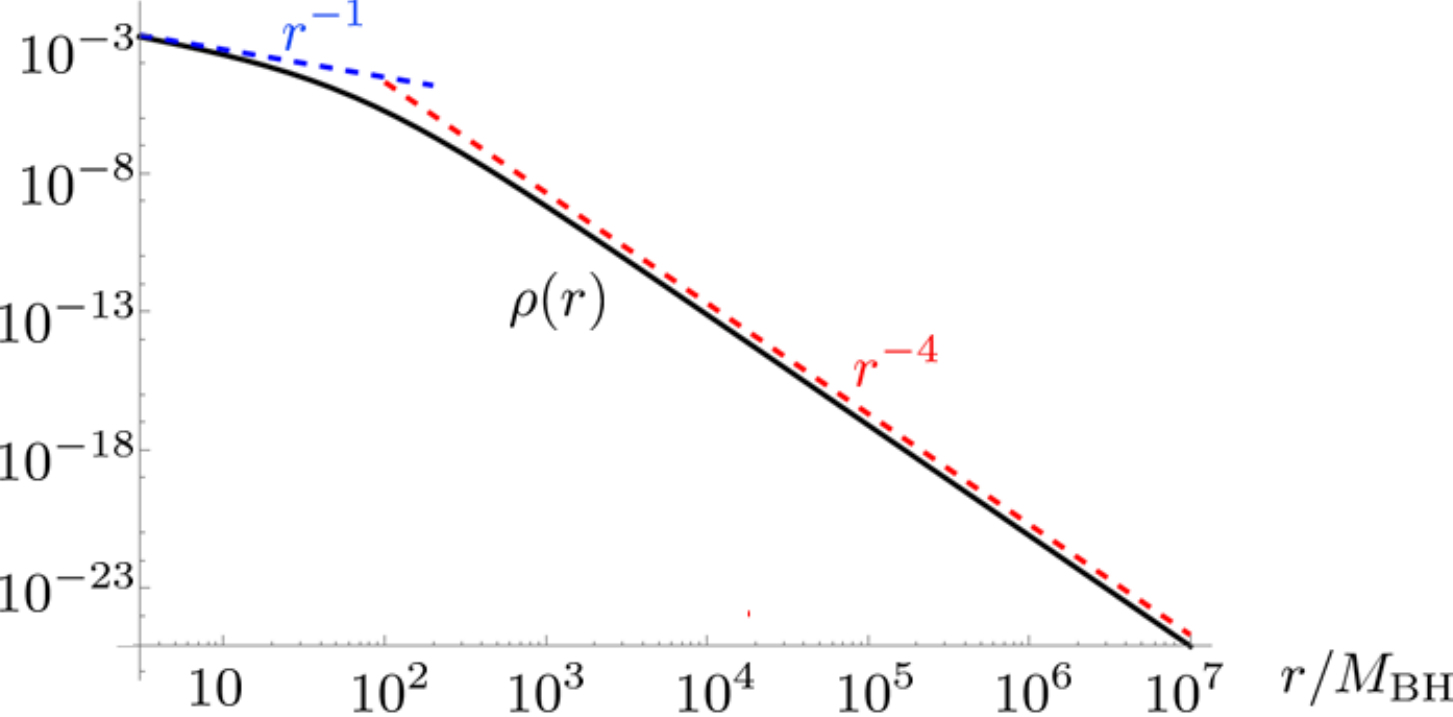}
\includegraphics[width=7cm]{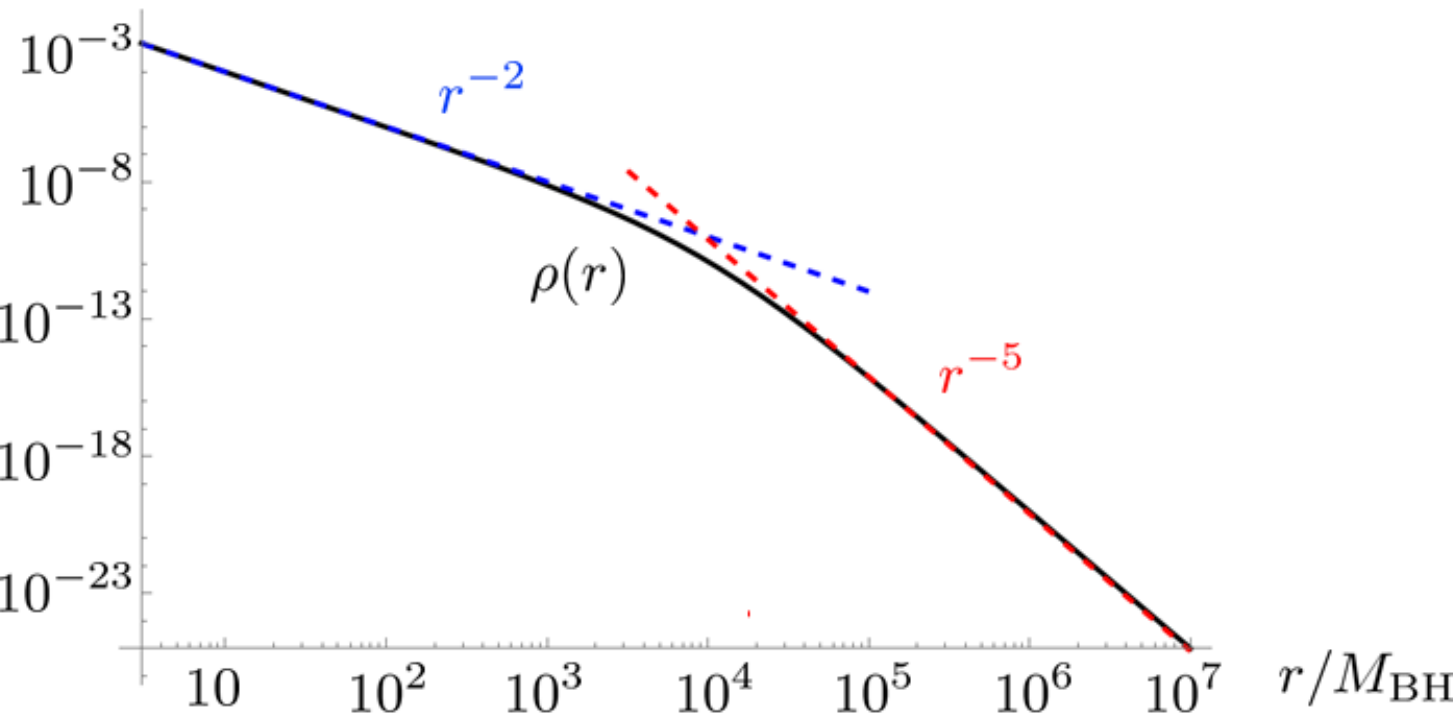}
\\
(a) \hskip 5cm (b)
\\
\includegraphics[width=7cm]{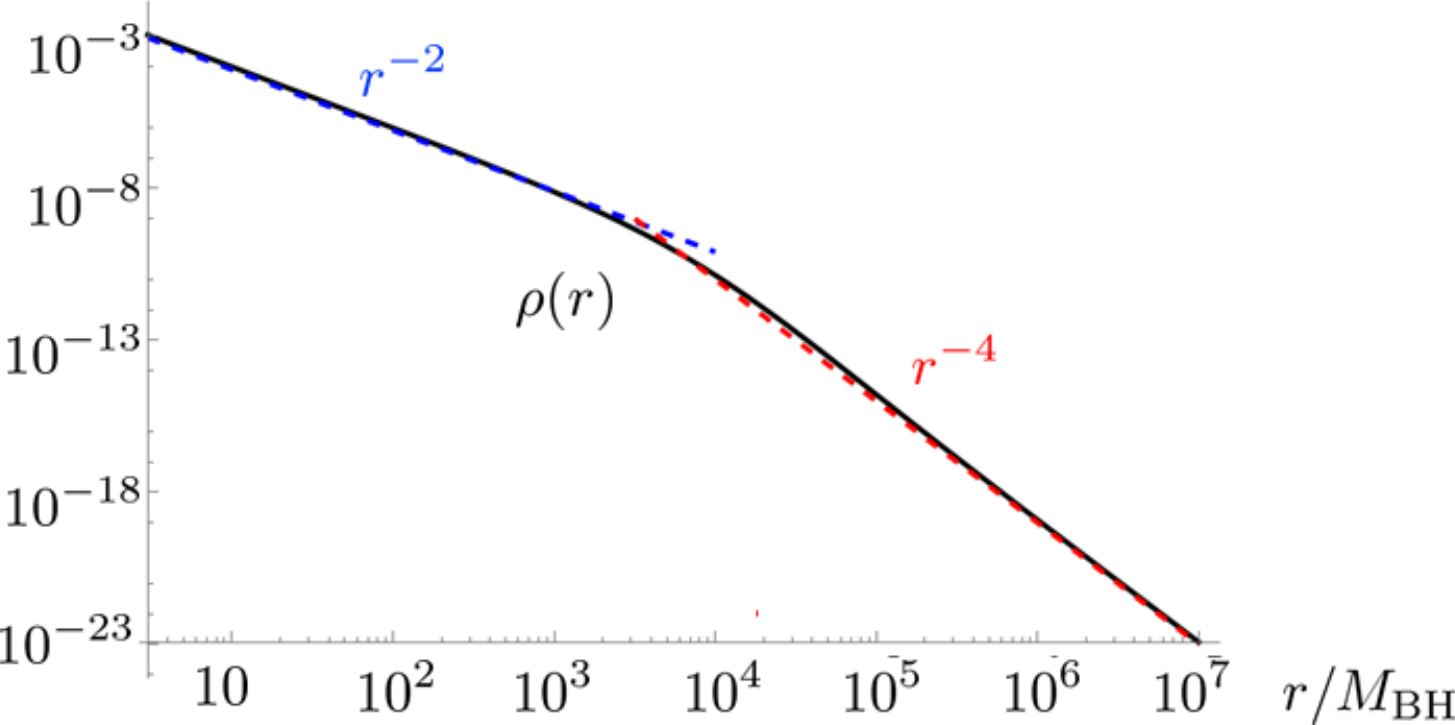}
\includegraphics[width=7cm]{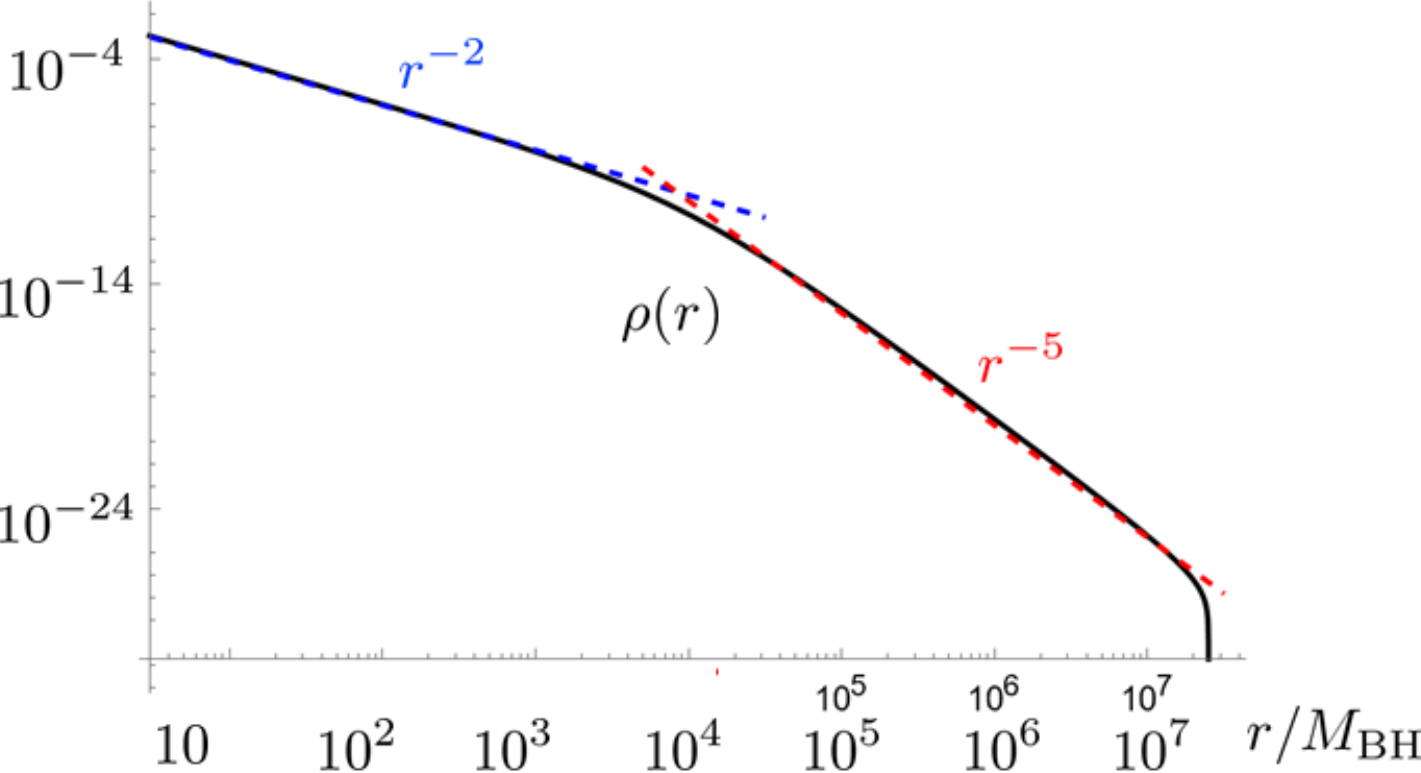}
\\
(c) \hskip 5cm  (d)
\caption{The density distributions of (a) Type A, (b)  Type B$_0$, (c)  Type B$_+$, and (d)  Type B$_-$. 
We choose $r_{\rm I}=4M_{\rm BH}$. 
}
\label{densityAB}
\end{figure}

\begin{figure}[ht]
\includegraphics[width=7cm]{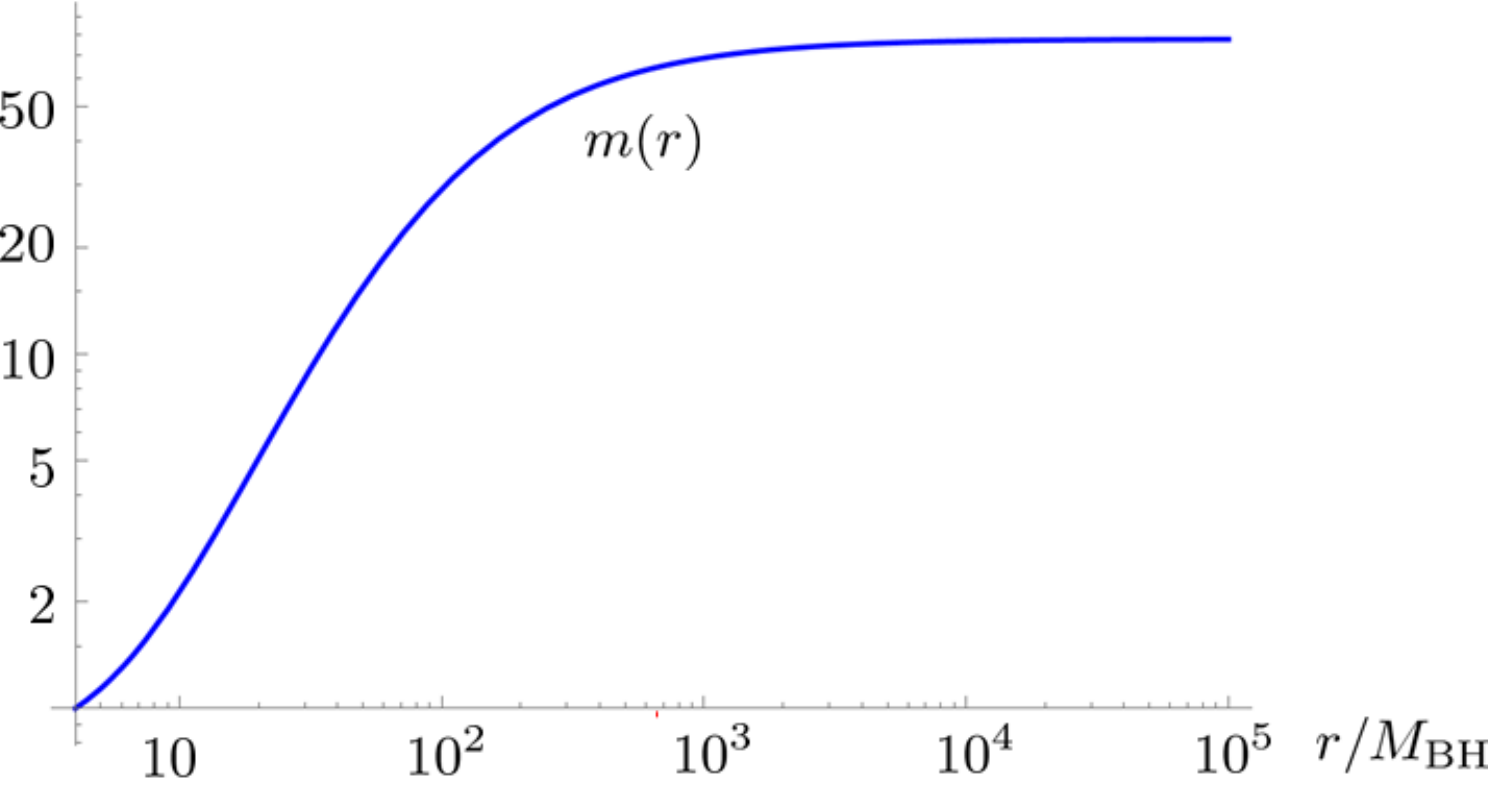}
\includegraphics[width=7cm]{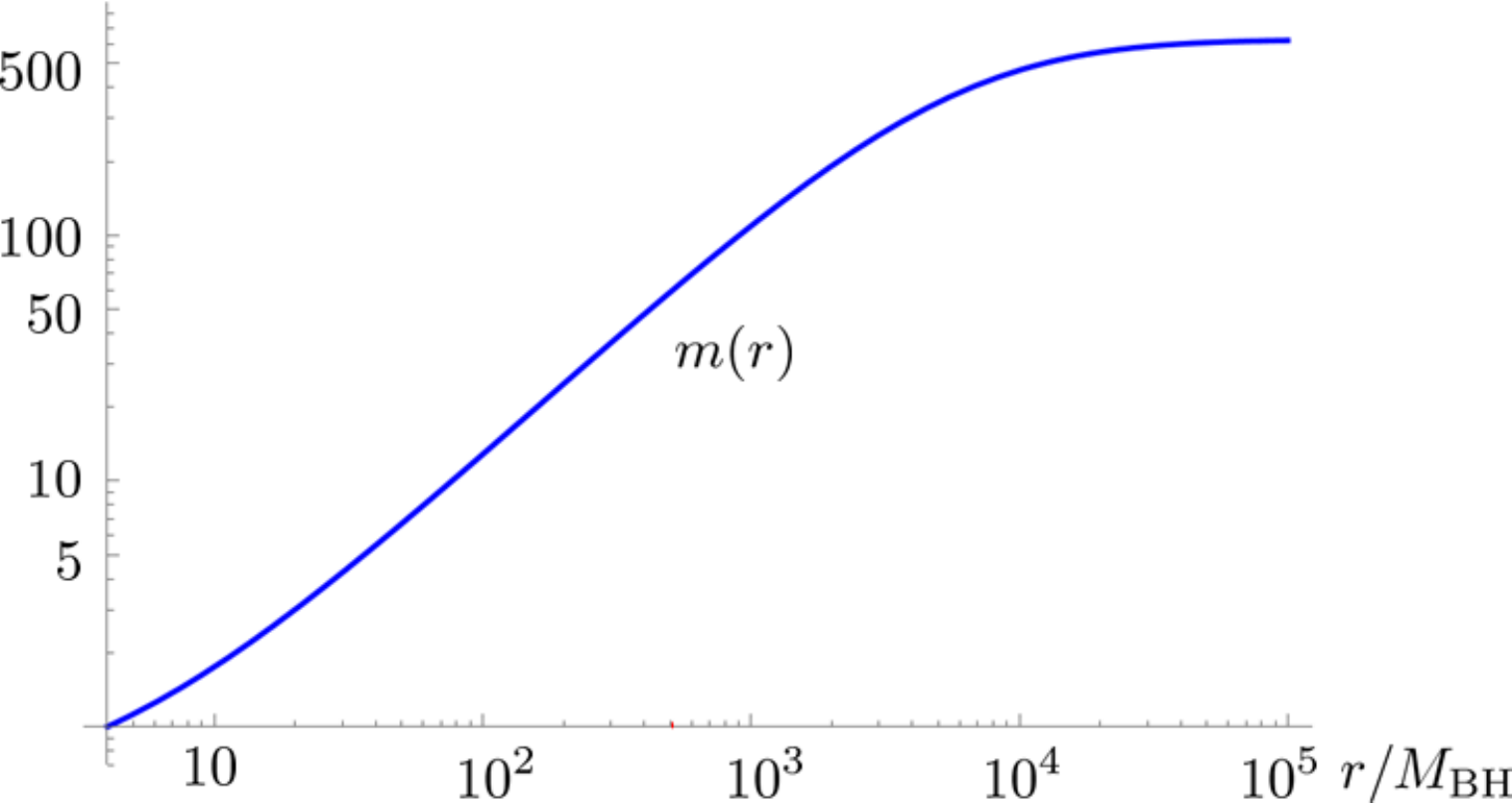}
\\
(a) \hskip 5cm (b)
\\
\includegraphics[width=7cm]{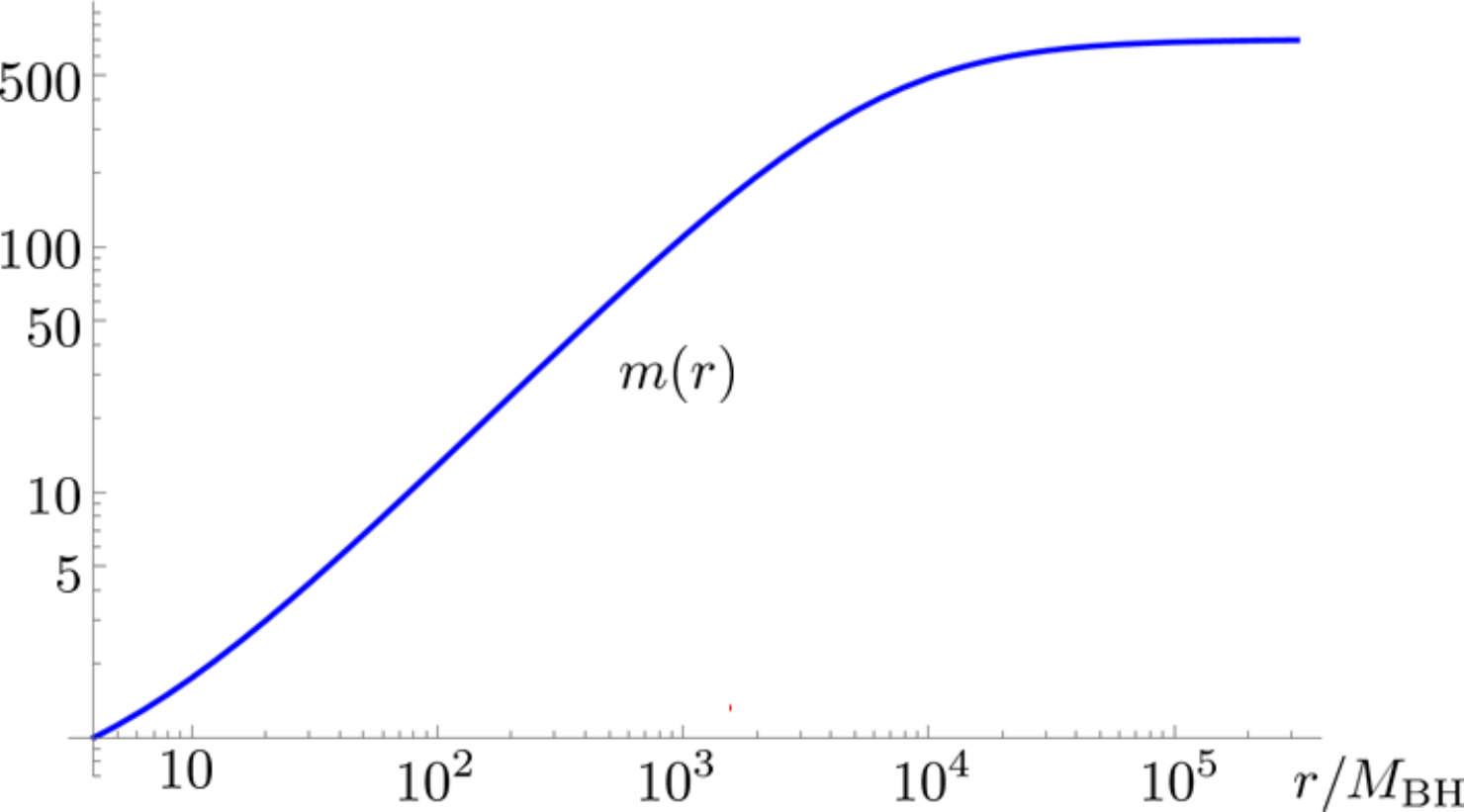}
\includegraphics[width=7cm]{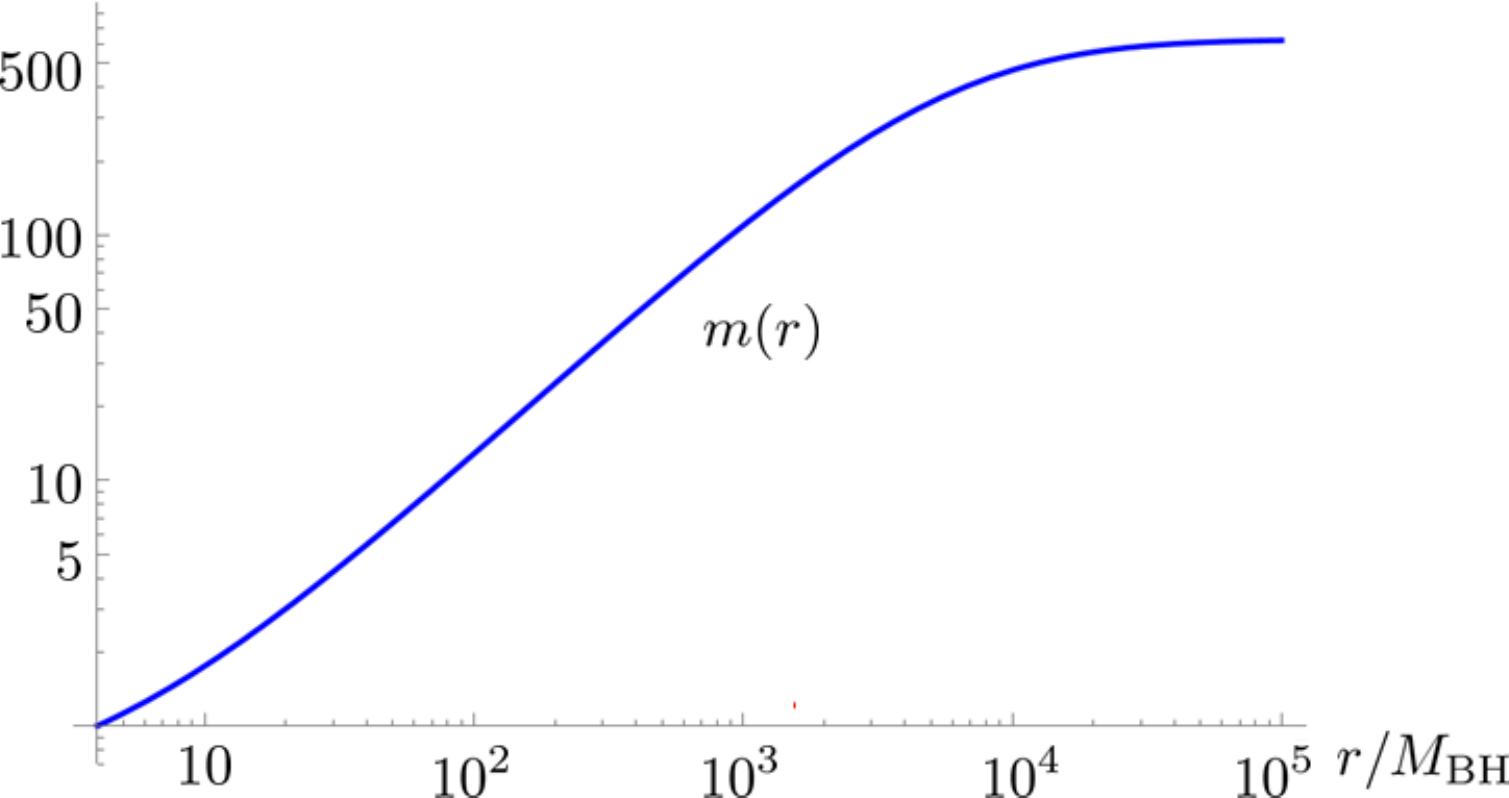}
\\
(c) \hskip 5cm  (d)
\caption{The mass functions of (a) Type A, (b)  Type B$_0$, (c)  Type B$_+$, and (d)  Type B$_-$. 
We choose $r_{\rm I}=4M_{\rm BH}$. 
}
\label{massAB}
\end{figure}

\begin{figure}[ht]
\includegraphics[width=7cm]{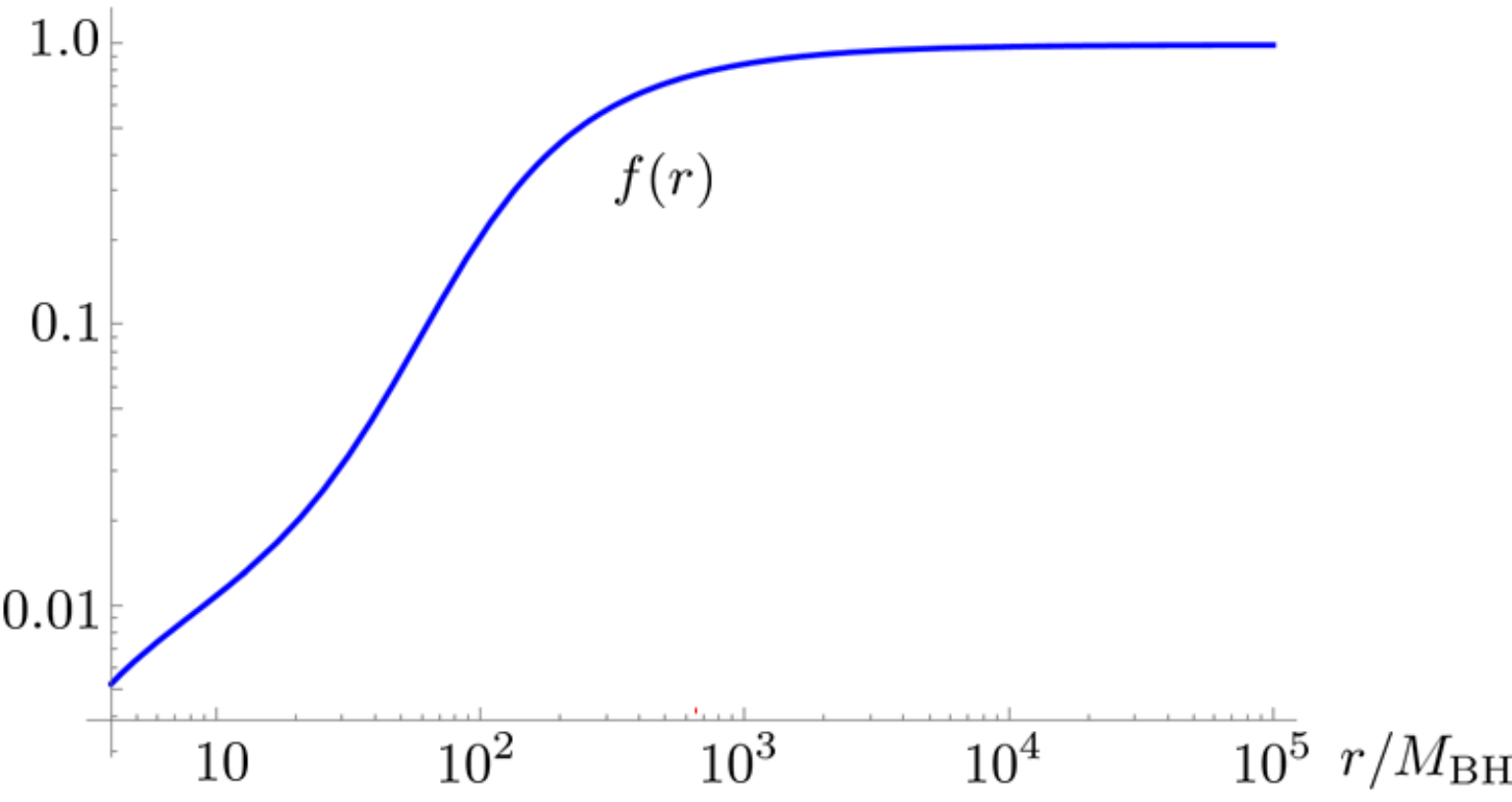}
\includegraphics[width=7cm]{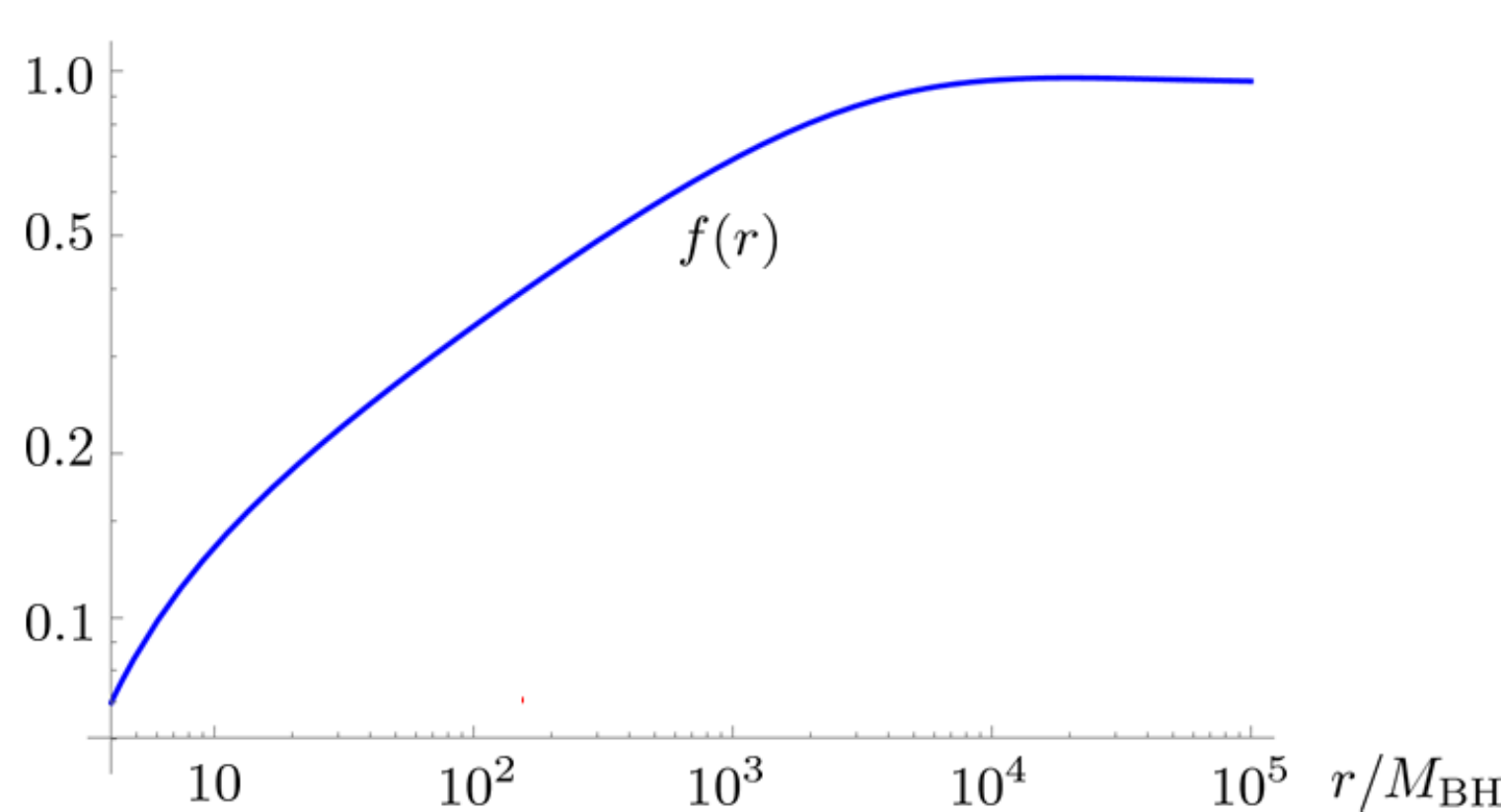}
\\
(a) \hskip 5cm (b)
\\
\includegraphics[width=7cm]{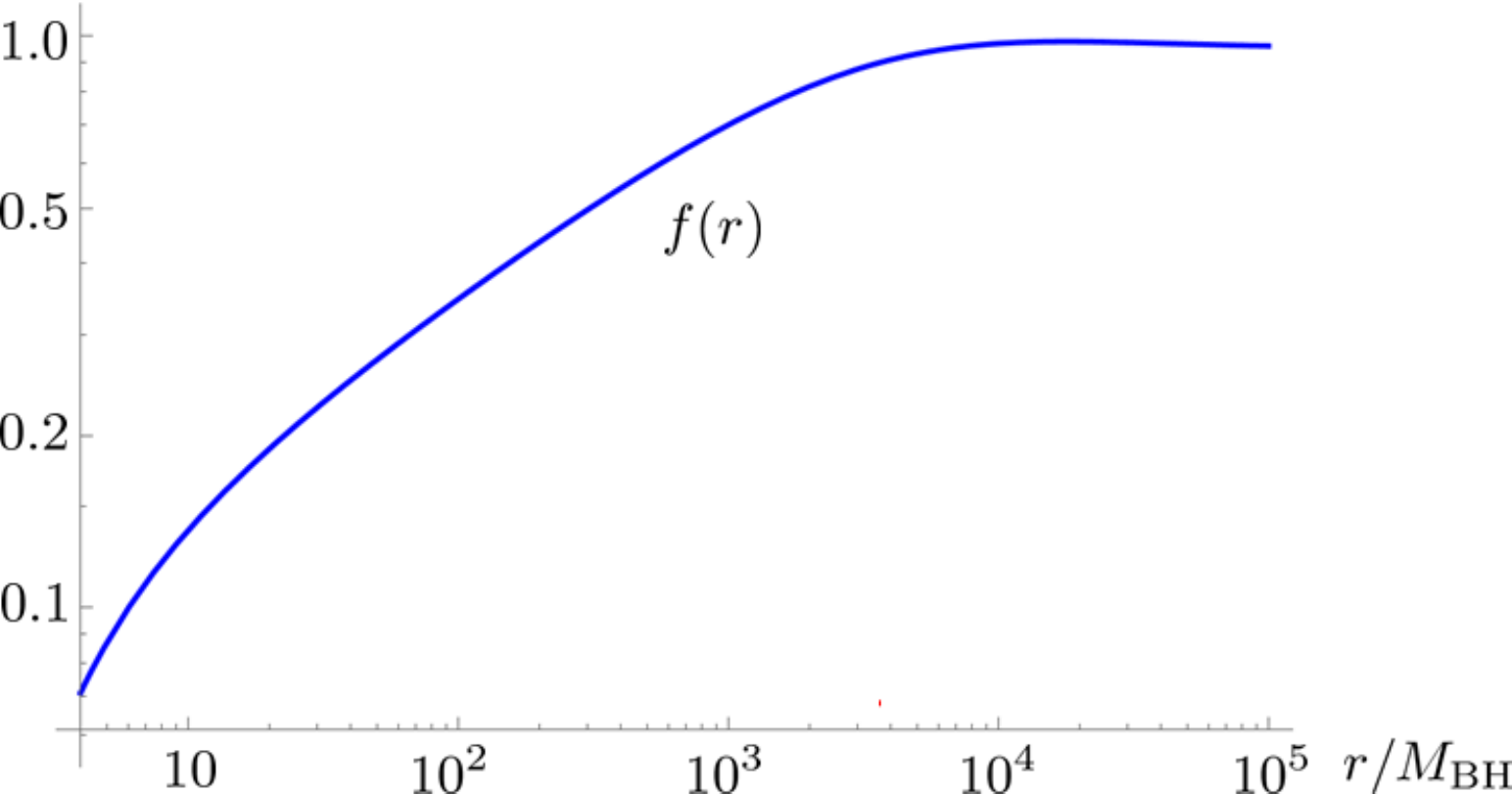}
\includegraphics[width=7cm]{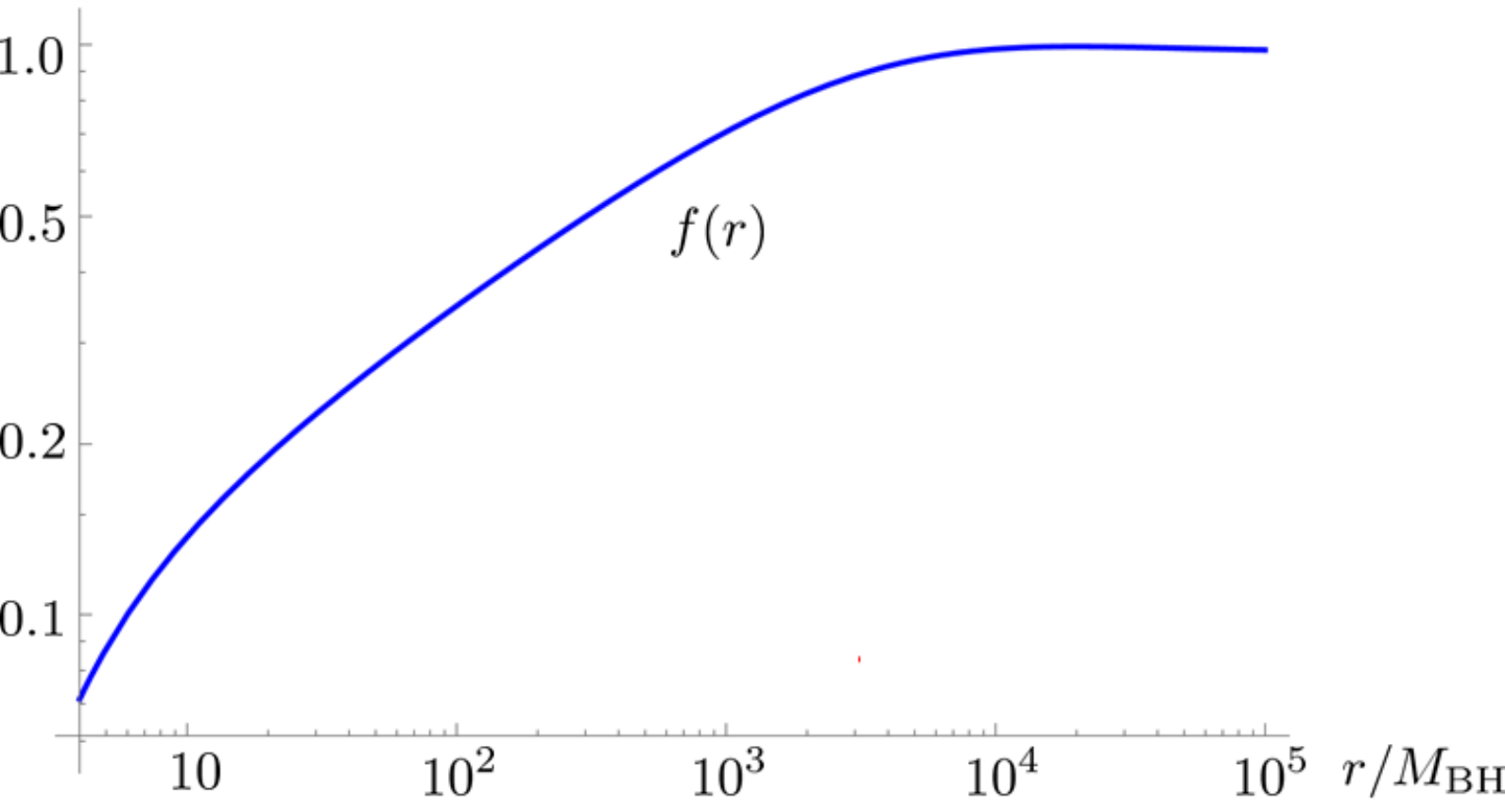}
\\
(c) \hskip 5cm  (d)
\caption{The metric functions $f$  of (a) Type A, (b)  Type B$_0$, (c)  Type B$_+$, and (d)  Type B$_-$. 
We choose $r_{\rm I}=4M_{\rm BH}$. 
}
\label{f_AB}
\end{figure}
\end{widetext}

In Fig. \ref{massAB}, we also show the 
mass functions  $m(r)$ for the same models.
 
As for the metric function $f(r)$, we find that the solution exists in whole range of $r_{\rm I}\leq r<\infty$
($r_{\rm I}\leq r<r_{\rm O}$ for Type B$_-$). The three roots 
(see Appendix for the meaning of the three roots) are
\begin{description}
  \item[{\normalfont Type A~:}]  $(1.555, 12.72\pm 60.88\,i$) ~~~~~~[case (2)] 
  \item[{\normalfont Type B$_0$:}] $( -12967., -5782., 1.33298$)\, [case (1-1)]
  \item[{\normalfont Type B$_+$:}]   $( -12696., -5906., 1.33299$)  [case (1-1)] 
  \item[{\normalfont Type B$_-$:}]  $( -12968.,  -5782., 1.33298$) [case (1-1)]
\end{description}

 Using these three roots, 
we  depict the 
 metric fubctions $f$ 
 for the same models in Fig. \ref{f_AB},. 

From these figures, we find three types B$_0$ and B$_\pm$ are very similar because the parameters we choose in B$_\pm$ are very close to those in  B$_0$.
If we choose them far from B$_0$, we find the different distributions. The behaviours of Type A are also very different from
those of Type B.

\newpage
  
\section{Summary and Remarks}
\label{summary}

Using the Einstein cluster model, 
we discuss general relativistic solutions describing possible BH environments. In particular, we show that nontrivial distributions can occur the (isolated) BH ISCO radius
($6M_{\rm BH}$). We have presented simple, realistic models with a specific galactic scale, as well as three toy models. The ISCO radius of the matter distribution is found to lie between the photon radius ($3M_{\rm BH}$) and 
the BH ISCO radius.

One of the  issues that we don't address is stability of these configurations. There  have been discussions regarding the stability of the Einstein clusters~\cite{GERALICO2012, acharyya2023modellingeinsteinclusterusing}. The Einstein cluster under our stability condition (\ref{stability}) 
appears to be metastable in the range of $3m(r)<r<6m(r)$, and it may become unstable when we consider a simple Einstein cluster star\cite{GERALICO2012}.
However there exists a BH inside the matter distribution in our models, and most of DM distribution satisfies the absolute stability condition ($r>6m(r)$) including the region outside the DM distribution.  Therefore, we must carefully reanalyze  the stability of such a system. 

The presence of matter in such a strong gravity region may be significant for gravitational wave observations, as well as for the DM search experiments, such as those involving particle annihilation or creation. We will leave these important studies, along
with the  stability analysis, for future work.

\begin{acknowledgments}
We thank Robin Diedrichs and Tomohiro Harada  for
useful discussions. 
We would like to acknowledge the Yukawa Institute for Theoretical
Physics at Kyoto University, where the present work was begun during
the YITP long-term workshop, 
Gravity and Cosmology 2024. 
KM would also thank Institute for Theoretical Physics and Cosmology,
Zhejiang University of Technology
 and Niels Bohr Institute/Niels Bohr International Academy, where this work was completed.
This work was supported in part by JSPS KAKENHI
Grant Number JP24K07058. 
V.C.\ is a Villum Investigator and a DNRF Chair.  
V.C. acknowledges financial support provided under the European Union’s H2020 ERC Advanced Grant “Black holes: gravitational engines of discovery” grant agreement no. Gravitas–101052587. 
Views and opinions expressed are however those of the author only and do not necessarily reflect those of the European Union or the European Research Council. Neither the European Union nor the granting authority can be held responsible for them.
This project has received funding from the European Union's Horizon 2020 research and innovation programme under the Marie Sklodowska-Curie grant agreement No 101007855 and No 101131233.
We acknowledge support by VILLUM Foundation (grant no. VIL37766) and the DNRF Chair program (grant no. DNRF162) by the Danish National Research Foundation.
A.W. is partially supported by the US NSF grant: PHY-2308845.
\end{acknowledgments}

\newpage
\appendix
\section{The metric function $f(r)$ in Model IV}
\label{Appendix}
In this appendix, we integrate the metric function $f(r)$.
 The metric function $f$ is given by
 \bea
 f=f_0\exp I(r)\,,
 \ena
 with 
  \beann
 I(r)
& =&
\int_{r_{\rm I}}^r dr {2(m_0+m_1 r+M r^2)\over r\left[r(r+r_*)^2-2(m_0+m_1 r+M r^2)\right]}
 \,.
 \enann

In order integrate it, we first consider the cubic equation  $p(r)\equiv r(r+r_*)^2-2(m_0+m_1 r+M r^2)=0$. It may contain 
three real roots, $r_1\,, r_2\,, r_3 (r_1\leq r_2\leq r_3)$
(case 1), or one 
real root $r_1$ and two complex conjugate roots $\xi \pm i \eta$ (case 2).
\\

\noindent
(1-1) three different real roots($r_1<r_2< r_3$)\\
In this case, we find 
\beann
 I(r)=\ln \left[r^a(r-r_1)^b(r-r_2)^c(r-r_3)^d\right]
 \,,
\enann
 where 
 \beann
a&=&-{2m_0\over r_1r_2r_3}\,,
 \\
b&=&{2[m_0+r_1(m_1-Mr_1)]\over r_1(r_2-r_1)(r_3-r_1)}\,,
 \\
c&=&-{2[m_0+r_2(m_1-Mr_2)]\over r_2(r_2-r_1)(r_3-r_2)}\,,
 \\
d&=&{2[m_0+r_3(m_1-Mr_3)]\over r_3(r_3-r_1)(r_3-r_2)}\,.
 \enann
 Hence the metric function is given by 
 \bea
 f(r)=f_0\, r^a(r-r_1)^b(r-r_2)^c(r-r_3)^d
 \,.
 \ena
 Since $a+b+c+d=0$, $f$ approaches a constant as 
 $r\rightarrow \infty$. 
 However, if at least one of $r_i\,(i=1,2,3)$ is larger than $r_I$, $f$ will diverges at that point. 
 The solution cannot be extended further. 
 \\
 
\noindent
(1-2) ($r_1=r_2< r_3$)\\
\beann
I(r)= a\ln r + b \ln (r - r_1) -{c\over (r - r_1)} + d \ln(r - r_3)\,,
\enann
where
 \beann
a&=&-{2m_0\over r_1^2r_3}\,,
 \\
b&=&{2[ m_0(2 r_1 -  r_3)+ (m_1   + M  r_3)r_1^2]\over r_1^2(r_3-r_1)^2}\,,
 \\
c&=&-{2[m_0+r_1(m_1+Mr_1)]\over r_1(r_3-r_1)}\,,
 \\
d&=&{2[m_0+r_3(m_1+Mr_3)]\over r_3(r_3-r_1)^2}\,.
 \enann
 Hence the metric function is given by 
 \bea
 f(r)=f_0\, r^a(r-r_1)^b(r-r_3)^d\exp
 \left[-{c\over (r - r_1)}\right]
 \,.~~~~
 \ena
 Since $a+b+d=0$, $f$ approaches a constant as 
 $r\rightarrow \infty$. 
 However, if at least one of $r_i\,(i=1,3)$ is larger than $r_I$, $f$ will diverges at that point. The solution cannot be extended further. 
\\

\noindent
(1-3) ($r_1<r_2= r_3$)\\
\beann
I(r)= a\ln r  + b \ln(r - r_1)+ c \ln (r - r_2)-{d\over (r - r_2)}\,,
\enann
where
 \beann
a&=&-{2m_0\over r_1r_2^2}\,,
 \\
b&=&{2[ m_0 + (m_1 + M r_1)r_1 ]
\over r_1(r_2-r_1)^2}\,,
 \\
c&=&-{2[m_0(- r_1 + 2  r_2) + m_1 r_2^2 + M r_1 r_2^2]\over r_2^2(r_2-r_1)^2}\,,
 \\
d&=&{2[m_0 + (m_1 + M r_2) r_2]\over r_2(r_2-r_1)}\,.
 \enann
Hence the metric function is given by 
 \bea
 f(r)=f_0\, r^a(r-r_1)^b(r-r_2)^c\exp
 \left[-{d\over (r - r_2)}\right]
 \,.~~~~~
 \ena
 Since $a+b+c=0$, $f$ approaches a constant as 
 $r\rightarrow \infty$. 
 However, if at least one of $r_i\,(i=1,2)$ is larger than $r_I$, $f$ will diverges at that point. The solution cannot be extended further. 
\\

\noindent
(1-4) ($r_1=r_2= r_3$)\\
\beann
I(r)= a\ln r  + b \ln(r - r_1)-{c\over (r - r_1)} -{d\over 2 (r - r_1)^2}\,,
\enann
where
 \beann
a&=&-{2m_0\over r_1^3}\,,
 \\
b&=&{2m_0 
\over r_1^3}\,,
 \\
c&=&{2[-m_0 +M r_1^2]
\over r_1^2}\,,
 \\
d&=&{2[m_0 + (m_1 + M r_1) r_1]\over r_1}\,.
 \enann
Hence the metric function is given by 
 \bea
 f(r)=f_0\, r^a(r-r_1)^b\exp
 \left[-{c\over (r - r_1)} -{d\over 2 (r - r_1)^2}\right]
 \,.
 \ena
 Since $a+b=0$, $f$ approaches a constant as 
 $r\rightarrow \infty$. 
 However, if  $r_1$ is larger than $r_I$, $f$ will diverges at that point. The solution cannot be extended further.

\noindent
(2) one real root and two complex conjugate roots \\
 In this case, the integrand is given as
 \beann
 {a\over r}+{b\over r-r_1}+{\gamma r+\delta \over r^2-2\xi r +\xi^2+\eta^2}
 \,,
 \enann
\begin{widetext}
where
 \beann
 a&=&-{2 m_0\over r_1 (\xi^2 + \eta^2)}\,,
 \\
 b&=&{2 (m_0 + m_1 r_1 + M r_1^2)\over r_1 (r_1^2 - 2 r_1 \xi + \xi^2 + \eta^2)}\,,
 \\
\gamma&=&-{2 \left[(-m_0  + M  (\xi^2 +  \eta^2))r_1+ 
2 m_0 \xi  + m_1 (\xi^2 + \eta^2) \right]\over (\xi^2 + \eta^2) (r_1^2 - 2r_1 \xi + \xi^2 + \eta^2)}\,,
 \\
\delta&=&{2 \left[-(2 m_0 \xi + m_1  (\xi^2+  \eta^2)) r_1
   + m_0(3  \xi^2-  \eta^2) + 2m_1\xi( \xi^2  + \eta^2 )
   + M (\xi ^2 +  \eta^2)^2 \right]
   \over (\xi^2 + \eta^2) (r_1^2 - 2 r_1 \xi +\xi^2 + \eta^2)}\,.
 \enann
We then find
\beann
I(r)=a\ln r+b\ln (r-r_1)+{\gamma\over 2}\ln (r^2 - 2 \xi r +\xi^2 + \eta^2)+{(\gamma \xi+\delta ) \over \eta}\arctan\left[{(r - \xi)\over \eta}\right]\,.
\enann
Hence the metric function is given by 
 \bea
 f(r)=f_0\, r^a(r-r_1)^b(r^2 - 2 \xi r +\xi^2 + \eta^2)^{\gamma/2}\exp\left[
 {(\gamma \xi+\delta ) \over \eta}\arctan\left[{(r - \xi)\over \eta}\right]\right]
 \,.
 \ena
 Since $a+b+\gamma=0$, $f$ approaches a constant as 
 $r\rightarrow \infty$. 
We also find $r_1$ is always smaller  than $r_I$ because
 $p(r_{\rm I})= r_{\rm I}(r_{\rm I}+r_*)^2-2(m_0+m_1r_{\rm I}+M r_{\rm I}^2)=(r_{\rm I}-2)(r_{\rm I}+r_*)^2>0$.
 Hence $f$ does not diverge anywhere. 
 
 ~~
 \vskip 1cm

\end{widetext}

\newpage
\bibliography{refer_DM}

\end{document}